\documentclass[11pt]{article}
\usepackage{authblk}

\usepackage[latin1]{inputenc}
\usepackage[T1]{fontenc}

\usepackage{authblk} 

\usepackage{amsmath}
\usepackage{amsfonts}
\usepackage{amssymb}

\usepackage[normalem]{ulem} 

\usepackage{amsmath,xcolor,ulem}
\usepackage{amsfonts}
\usepackage{amssymb,todonotes}
\usepackage{amsthm}
\usepackage{graphicx}
\usepackage{marginnote}

\usepackage[colorlinks=true, allcolors=blue]{hyperref}

\usepackage[top=2.8cm,bottom=2.8cm,left=2.5cm,right=2.5cm]{geometry}
\usepackage{float}
\usepackage[algoruled]{algorithm2e}
\usepackage[font=footnotesize,width=\linewidth]{caption}

\usepackage{graphicx}
\usepackage{amssymb}
\usepackage{amsthm}
\usepackage{amsmath}
\usepackage{algorithmic}
\usepackage{subcaption}
\usepackage{caption}
\usepackage{float}
\usepackage{mathrsfs} 
\usepackage{diagbox}
\usepackage{verbatim}

\newtheorem{definition}{Definition}
\newtheorem{theorem}{Theorem}
\newtheorem{remark}{Remark}
\newtheorem{proposition}{Proposition}
\newtheorem{example}{Example}

\title{Port-Hamiltonian Neural Networks: From Theory to Simulation of Interconnected Stochastic Systems}

\author{Luca Di Persio\footnote{ \href{mailto:luca.dipersio@univr.it}{ luca.dipersio@univr.it}} ,
Matthias Ehrhardt\footnote{ \href{mailto:ehrhardt@uni-wuppertal.de}{ehrhardt@uni-wuppertal.de}} ,
Youness Outaleb\footnote{Corresponding author, \href{mailto:youness.outaleb@unitn.it}{youness.outaleb@unitn.it}} ,
Sofia Rizzotto\footnote{ \href{mailto:sofia.rizzotto@studenti.univr.it}{sofia.rizzotto@studenti.univr.it}} 
}

\affil{Department of Computer Science -- College of  Mathematics\\ University of Verona, Italy}
\affil{IMACM, School of Mathematics and Natural Sciences, \\ University of Wuppertal, Germany}

\begin{document}
\maketitle

\begin{tikzpicture}[remember picture,overlay]
	\node[anchor=north east,inner sep=20pt] at (current page.north east)
	{\includegraphics[scale=0.2]{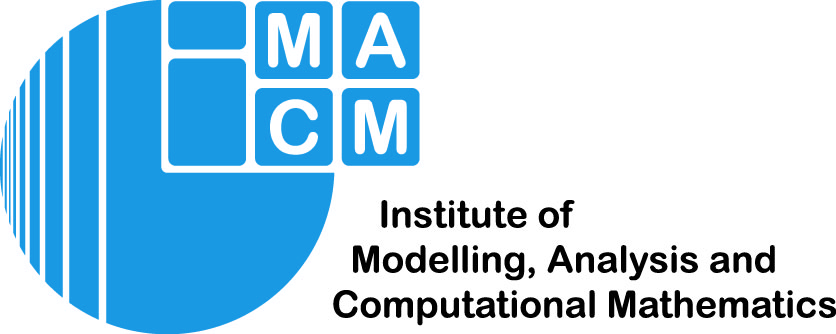}};
\end{tikzpicture}

\begin{abstract}
This work introduces a new framework integrating port-Hamiltonian systems (PHS) and neural network architectures. This framework bridges the gap between deterministic and stochastic modeling of complex dynamical systems. We introduce new mathematical formulations and computational methods that expand the geometric structure of PHS to account for uncertainty, environmental noise, and random perturbations. Building on these advances, we introduce stochastic port-Hamiltonian neural networks (pHNNs), which facilitate the accurate learning and prediction of non-autonomous and interconnected stochastic systems. \\
Our proposed framework generalizes passivity concepts to the stochastic regime, ensuring stability while maintaining the system's energy-consistent structure. Extensive simulations, including those involving damped mass-spring systems, Duffing oscillators, and robotic control tasks, demonstrate the capability of pHNNs to capture complex dynamics with high fidelity, even under noise and uncertainty. This unified approach establishes a foundation for the robust, data-driven modeling and control of nonlinear stochastic systems.
\end{abstract}

\begin{minipage}{0.9\linewidth}

\medskip

\end{minipage}

%%%%%%%%%%%%%%%%%%%%%%%%%%%%%%%% Intro
\section{Introduction}\label{sec1}
This research introduces a pioneering approach that combines \textit{port-Hamiltonian systems} (PHS) with neural network architectures. 
The focus is on transitioning from deterministic frameworks to stochastic models. 
Through an in-depth analysis of sophisticated mathematical formulations, we seek to enhance our comprehension of dynamical systems operating under uncertainty, encompassing a variety of intricate interactions and the potential for measurement errors.

PHS are defined as a synthesis of port modeling and geometric Hamiltonian dynamics, focusing on the Dirac structure. The Dirac structure generalizes traditional Poisson and pre-symplectic frameworks, enabling a nuanced representation of energetic topology within dynamical systems.
PHS's stochastic adaptation incorporates uncertainties, such as inherent noise and environmental variabilities, directly into the system's ports.
In this context, noise is not merely an external disturbance but an integral component of the system`s behavior.

In our work, we offer a coordinate-free geometric formulation of deterministic PHS. This formulation employs generalized Poisson brackets and Hamiltonian functions to describe the underlying dynamics comprehensively.
We seamlessly transition these theoretical constructs into local equations characterized by Jacobians and vector fields, providing a clear mathematical foundation. Furthermore, we extend this formulation to implicit systems with algebraic constraints by modeling energy storage and exchange using flow and effort variables and power ports.

The Dirac structure elucidates power-conserving interconnections within the system, which are essential for maintaining the principle of passivity. This property is crucial to stability in control applications. Building on this theoretical groundwork, we derive stochastic \textit{port-Hamiltonian neural networks} (pHNNs) by incorporating noise within the port structure.
This innovative framework is based on the methodology established by Cordoni, Di Persio, and Muradore  \cite{Cordoni22}, which adeptly integrates random perturbations to mitigate the impacts of measurement noise, parameter uncertainties, and environmental interactions, hence enriching the model's robustness.

The paper is organized as follows: 
Section~\ref{sec1} introduces key concepts and the various types of PHS, including discrete forms and their integration with neural networks.
Basic concepts and definitions for PHS are introduced in Section~\ref{sec1b}.
Section~\ref{sec2} extends PHS to include random elements for modeling systems with uncertainties. 
Section~\ref{sec3} generalizes the concept of \textit{passivity} to stochastic systems by defining strong and weak passivity, as well as providing criteria for passivity in \textit{stochastic port-Hamiltonian systems} (SPHS). 
Section~\ref{sec4} applies and develops stochastic PHS in different contexts. 
Section~\ref{sec5} presents simulation results of pHNNs on damped mass-spring and chaotic Duffing systems.
Section 6 concludes with a summary and discusses how we adapted Colonius and Gr\"une's neural network-based controller design method for systems with stochastic dynamics \cite{Colonius02}.

\section{Preliminaries on Port-Hamiltonian and Stochastic Models}\label{sec1b}
% \section{Theoretical Background}\label{sec1b}
% Basic Concepts and Definitions
%%%%%%%%%%%%%%%%%%%%%%%%%%%%%%%%%%%%%%% Section 1.1
\subsection{Basic Concepts}\label{sec11}
Cordoni, Di Persio, and Muradore \cite{Cordoni22} describe an \textit{input-state-output} (I-S-O) deterministic PHS using a geometric, coordinate-free formulation in terms of Poisson brackets:
\begin{equation}\label{a}
    \begin{cases}
        \dot{x} &= X_H(x)+\sum\limits_{i=1}^m u_i X_{H_{g_i}}(x),\\
            y_i &= \{H,H_{g_i}\}.
    \end{cases}
\end{equation}
This is called an \textit{(explicit) input-state-output port-Hamiltonian system} (PHS) on a Poisson manifold $(X,\{\cdot,\cdot\})$ with a Hamiltonian function $H\in C^\infty(\mathcal{X})$, $x\in\mathbb{R}^n$, the $i$-th input $u_i\in U$, the $i$-th output $y_i\in U^*$,
and the Hamiltonian vector field $X_{H_{g_i}}$ associated with the Hamiltonian $H_{g_i}$. 
In local coordinates, the previous system \eqref{a} reads
\begin{equation}\label{a2}
    \begin{cases}
        \dot{x} &= J(x)\partial_x H(x) + \sum\limits_{i=1}^m u_i g_i(x),\\
           y_i  &= g_i^\top(x) \partial_xH.
    \end{cases}
\end{equation}
Moreover, given $X_H^L(\cdot):=[\cdot,H]_L$, we can define the \textit{(explicit) input-state-output port-Hamiltonian system with dissipation} as follows:
\begin{equation}\label{b}
    \begin{cases}
        \dot{x} &= X_H^L(x) + \sum\limits_{i=1}^mu_iH_{g_i}(x),\\
        y_i &= [H,H_{g_i}],
    \end{cases}
\end{equation}
and in local coordinates, this system \eqref{b} reads
\begin{equation}\label{b2}
    \begin{cases}
        \dot{x} &= \bigl( J(x)-R(x) \bigr) \partial_x H(x) 
                         + \sum\limits_{i=1}^m u_i g_i(x),\\
            y_i &= g_i^\top(x) \partial_xH(x),
    \end{cases}
\end{equation}
where $R(x):=(g^R(x))^\top \Tilde{R}(x) g^R(x)$. 

Consider a physical system consisting of energy-storing elements, energy-dissipating elements, and power ports.
These elements are connected by power-preserving links, which can only transfer energy and cannot produce it.
Such a system can be described by extending the port-Hamiltonian system framework to implicit systems, i.e., systems with algebraic constraints.
Given a state space $\mathcal{X}$ (a smooth manifold whose elements represent the energy stored in the system), a vector space of flow variables $\mathcal{V}$ and its dual space of effort variables $\mathcal{V}^*$ (representing the power ports), a geometric Dirac structure, $\mathcal{D}$, and a Hamiltonian function, $\mathcal{H}$, representing the total energy of the system in a given state, we can define an \textit{implicit port-Hamiltonian system} corresponding to $(\mathcal{X},\mathcal{V},\mathcal{D},\mathcal{H})$ as
\begin{equation}
    v = -\dot{x}\quad\text{and}\quad v^*=\frac{\partial\mathcal{H}}{\partial x}(x)\,,
\end{equation}
implying the system is defined by
\begin{equation}
    \Bigl( -\dot{x},\frac{\partial H}{\partial x}(x),f,e \Bigl)\in\mathcal{D}(x).
\end{equation}
The Dirac structure discussed here describes the internal interconnection behavior of a port-Hamilto\-nian system and provides a mathematical framework for understanding how its components interact.
A key attribute of Dirac structures is their power-conserving composition yields another valid Dirac structure.
This property is crucial because it implies that any interconnection of port-Hamiltonian systems that preserves power will also result in a valid system that maintains the underlying principle of energy conservation. 
Specifically, when combining these systems, the overall Dirac structure is constructed by integrating the individual structures, and the total Hamiltonian of the interconnected system is expressed as the sum of the Hamiltonians of its components.

Furthermore, a fundamental characteristic of these systems is the concept of \textit{passivity}, which states that the total energy supplied to the system must equal or exceed the energy released, assuming no losses due to friction, resistance, or other dissipative effects. 
This principle of passivity is essential for ensuring stability in various control applications because it reveals how a system behaves in response to energy inputs and outputs. 
Passivity naturally arises from the underlying Dirac structure and is closely tied to the energy-dissipation relationship inherent to port-Hamiltonian systems. 
This relationship guarantees that energy is neither created nor destroyed within the system, only shifted between different forms and components.
Ultimately, this relationship facilitates robust and stable control strategies.

In summary, understanding the dynamics and stability of
%port-Hamiltonian systems
PHSs require grasping the interplay between the Dirac structure, power conservation, and passivity.
This interplay allows us to design and analyze complex, interconnected systems for various engineering and physical applications.

%%% This text is double basically
%The Dirac structure described above characterizes the system's internal interconnection behavior. A key property of Dirac structures is that their power-conserving composition yields another valid Dirac structure. Hence, any power-preserving interconnection of port-Hamiltonian systems forms a valid port-Hamiltonian system. In such cases, the overall Dirac structure is constructed from the composition of the individual structures, and the total Hamiltonian is the sum of the component Hamiltonians.

%%% This text is double basically
%Another fundamental property of these systems is \textit{passivity}, which ensures that the total energy supplied to the system is never less than the energy it releases, assuming there are no losses. Passivity plays a central role in guaranteeing stability in control applications. Passivity arises naturally from the Dirac structure and the inherent energy-dissipation relationship within PHSs.

%%%%%%%%%%%%%%%%%%%%%%%%%%%%%%%%%%%%%%% Section 1.2
\subsection{Discrete Systems}\label{sec12}
Until now, we have only considered port-Hamiltonian systems in continuous time.
However, it is crucial to determine if their fundamental properties, such as energy conservation, are maintained during time discretization. 
As Viswanath, Clemente-Gallardo, and van der Schaft \cite{Talasila06} demonstrated, Hamiltonian systems can be discretized to preserve energy-conserving behavior.
Discrete port-Hamiltonian systems can be formed by discretizing continuous models or formulating them directly in discrete time. 
In the latter case, Poisson brackets % can be used because they 
are used to retain the critical structural properties necessary to preserve the Hamiltonian framework in the discrete domain: skew symmetry, bilinearity, and a modified Leibniz rule.

\begin{figure}[htbp]
\centering
\setlength{\unitlength}{0.8cm}
\includegraphics[width=.90\textwidth]{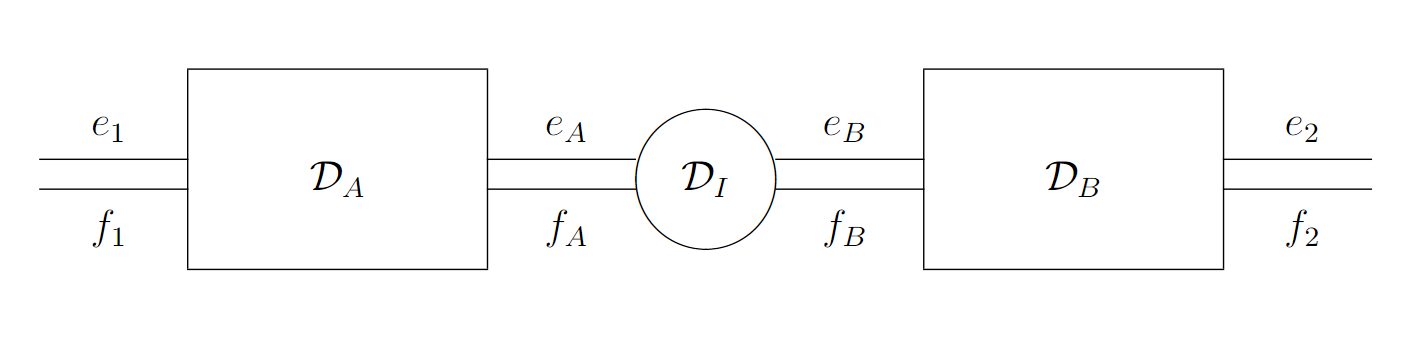}
\caption{Interconnection of two port-Hamiltonian systems.}\label{figure12}
\end{figure}

First, we must define the discrete Dirac structure.
To do so, we denote a space of discrete vector fields by $\mathfrak{X}(A)$ and a space of discrete 1-forms by $\Lambda^1(A)$.
Then, a \textit{generalized Dirac structure} on an $n$-dimensional discrete manifold is an $n$-dimensional linear subspace $\mathcal{D}\subset\mathfrak{X}(A)\times\Lambda^1(A)$ such that $\mathcal{D}=\mathcal{D}^\top$ with
\begin{equation*}
    \mathcal{D}^\top=\bigl\{(Y,\beta)\in\mathfrak{X}(A)\times\Lambda^1(A)\quad
    \text{where}\quad\langle \alpha,X\rangle+\langle\beta,Y\rangle=0,\quad
    \forall\,(X,\alpha)\in\mathcal{D}\bigr\},
\end{equation*}
where $\langle\cdot,\cdot\rangle$ is the pairing between $\mathbb{F}^n$ and $\mathbb{F}^{n*}$.
Consider the configuration of the effort-flow pairs shown in Figure~\ref{figure12}.
Let $\mathcal{F}_i$ denote the space of flow and $\mathcal{D}_i$ the space of effort of the Dirac structure for systems $i=A,B$. 
The \textit{interconnection between the two Dirac structures} $\mathcal{D}_A$ and $\mathcal{D}_B$ can then be defined as
\begin{equation}
    \begin{split}
        \mathcal{D}_A\circ\mathcal{D}_B &:= \big\{(f_1,e_1,f_2,e_2)\in\mathcal{F}_1\times\mathcal{F}_1^*\times\mathcal{F}_2\times\mathcal{F}_2^*\quad\text{such that}\\ 
        &\exists(f,e)\in\mathcal{F}\times\mathcal{F}^*\quad\text{with}\quad
        (f_1,e_1,f,e)\in\mathcal{D}_A\quad\text{and}\quad(-f,e,f_2,e_2)\in\mathcal{D}_B\big\}.
        \end{split}
\end{equation}
Then $\mathcal{D}_A\circ\mathcal{D}_B$ is a Dirac structure.

%%%%%%%%%%%%%%%%%%%%%%%%%%%%%%
%%%%%%%%%%%%%%%%%%%%%%%%%%%%%%%%

\begin{definition}[Implicit discrete port-Hamiltonian system \cite{Talasila06}]
    Let $H\colon\mathcal{Z}\to\mathbb{F}$ be a discrete Hamiltonian, $\mathcal{F}_P$ be the space of external flows $f$, $\mathcal{E}_P=\mathcal{F}_P^*$ be the space of external effort $e$, and $\mathcal{D}$ be the Dirac structure depending only on the coordinate $z$. 
    Then, the \textit{implicit discrete port-Hamiltonian system} is defined as
    \begin{equation}
        \Bigl( -\frac{\Delta z}{\Delta t},f,\Game_zH(z),e \Bigl) \,\in\mathcal{D}(z).
    \end{equation}
\end{definition}
A system may consist of continuous and discrete port-Hamiltonian components encountered in various applications.
To effectively manage these hybrid systems, Viswanath, Clemente-Gallardo, and van der Schaft \cite{Talasila05}  proposed a methodology that reinterprets energy-conserving interconnections at discrete sampling points as interconnections incorporating an external flow source.
This transformation ensures energy conservation throughout sampling intervals, preserving the passive and port-Hamiltonian characteristics of the system.

This approach allows for integrating discrete computational models with continuous physical systems without compromising overall stability or passivity. 
However, it is noteworthy that this setup does not guarantee exact energy conservation because of the discrete sampling intervals.

To address the potential energy conservation issue, Kotyczka and Lefevre \cite{Kotyczka19} introduced a discrete-time Dirac structure and a discrete-time port-Hamiltonian framework. 
Their methodology uses symplectic integration techniques, specifically collocation methods, to approximate the energy balance observed continuously. 
Their approach maintains the fidelity of the system's dynamics to the original continuous-time port-Hamiltonian representation by ensuring structural consistency over discrete time steps, allowing for the effective integration of numerical algorithms.

%%%%%%%%%%%%%%%%%%%%%%%%%%%%%%%%%%%%%%%%%%%%%%%%%%%%%%%%%%%%% Section 1.3
%\subsection{Port-Hamiltonian Neural Networks}\label{sec13}

%%%%%%%%%%%%%%%%%%%%%%%%%%%%%%%%%%%%%%%%%%%%%%%%%%%%% Section 2
\section{Stochastic port-Hamiltonian systems}\label{sec2}
% Things become more subtle when moving to a stochastic setting.
In this section, we move to a stochastic setting.
Random perturbations cause system trajectories to become stochastic processes. 
The deterministic duality pairing previously used to define Dirac structures must now be interpreted using \textit{stochastic integrals}; moreover, it also necessitates a generalized definition of orthogonality that remains consistent with the energetic interpretation under stochastic dynamics.

Let $\bigl(\Omega,\mathcal{F},(\mathcal{F}_t)_{t\in\mathbb{R}_+},\mathbb{P}\bigl)$ be a complete probability space. 
Denote the Stratonovich integral by $\delta Z$ and the It\^{o} integral by $dZ$ along the semimartingale $Z$.
As previously discussed, the stochastic port-Hamiltonian framework incorporates randomness by modeling each system component as a semimartingale.
 Using Stratonovich calculus aligns with the geometric structure of Dirac formulations. 
 However, for analytical convenience and to leverage the probabilistic tools of stochastic analysis,  this formulation can equivalently be expressed in terms of It\^{o} calculus.
Consider the system
\begin{equation}\label{lcsphs}
    \begin{cases}
        \delta X_t &= \bigl(J(X_t)-R(X_t) \bigr)\, \partial_x H(X_t)\,\delta Z_t 
        + g(X_t)u\,\delta Z^g_t + \xi(X_t)\,\delta Z^N_t,\\
               y_t &= g^\top(X_t)\,\partial_x H(X_t),
    \end{cases}
\end{equation}
where $R(x):=(g^R(x))^\top\Tilde{R}(x)g^R(x)$, $W$ denotes a Brownian motion, and $Z$, $Z^g$, and $Z^N$ are semimartingales. 
Then \eqref{lcsphs} describes the stochastic PHS in local coordinates, and the obtained framework can be further extended to scenarios in which noise enters the system by generalizing the concept of a Dirac structure. 
Noise can enter the system as a stochastic external field or as random perturbations affecting any of the system's connected ports.

%%%%%% Motivation for Orthogonal Complement and Generalized Dirac Structure}
% When modeling physical systems, particularly those governed by port-Hamiltonian formulations, a primary objective is to preserve the \textit{energetic structure} of the system, even when considering interconnection, control, or discretization. When randomness is introduced through environmental perturbations, measurement noise, or inherent stochasticity - the energetic structure must still be respected to maintain meaningful dynamics.

% We rely on the concept of a \textit{Dirac structure} to express this. A Dirac structure captures the power-conserving relationships between \textit{flows} and \textit{efforts} in a system. In a deterministic setting, Dirac structures ensure that pairing these variables does not create or destroy energy, thereby preserving \textit{passivity} and enabling stable control design.

%When moving to a \emph{stochastic setting}, things become more subtle. Random perturbations mean that system trajectories become stochastic processes, and the deterministic duality pairing used to define Dirac structures must now be interpreted over \emph{stochastic integrals}. This leads to the need for a generalized definition of orthogonality that remains consistent with the energetic interpretation under stochastic dynamics.

Next, we introduce the \emph{orthogonal complement} $\mathcal{D}^\perp$, consisting of all pairs $(\delta X_t, \sigma)$ that satisfy an energy-balance relation (expressed using \emph{Stratonovich integrals}) concerning all elements of a given subbundle $\mathcal{D}$. 
The condition ensures no net energy is produced or lost by pairing these elements over time.

%%%%%%%%%%%%%%%%%%%%%%%%%%%%%%%%%%%%%%%%%%%%%%%%%
\begin{definition}[Orthogonal complement, \cite{Cordoni22}]\label{def:oc}
    Given a manifold $\mathcal{X}$, $I\subset\mathbb{R}_+$, a bundle $\mathcal{D}\subset T\mathcal{X}\oplus T^*\mathcal{X}$, a differential 1-form $\sigma$ on $\mathcal{X}$ and an integral curve $X\colon I\to\mathcal{X}$ of a Stratonovich vector field $\delta X_t$, the \textit{orthogonal complement} of $\mathcal{D}$ is 
    \begin{equation}
        \mathcal{D}^\perp=\bigl\{(\delta X_t,\sigma)\subset T\mathcal{X}\oplus T^*\mathcal{X}\colon\int_0^t\langle \sigma,\delta \bar{X}_s\rangle + \int_0^t\langle \bar{\sigma},\delta X_s\rangle =0,\; 
        \forall\,(\delta\bar{X}_t,\bar{\sigma})\in\mathcal{D},\ t\in I\bigr\}
    \end{equation}
\end{definition}

From this, we define a \emph{generalized stochastic Dirac structure} as a subbundle $\mathcal{D} \subset T\mathcal{X} \oplus T^*\mathcal{X}$ that equals its orthogonal complement: $\mathcal{D} = \mathcal{D}^\perp$. 
This condition ensures that $\mathcal{D}$ is both \emph{isotropic} (energetically neutral) and \emph{maximal} (it includes all such neutral pairs), thereby generalizing the classical concept of a Dirac structure to systems evolving under stochastic dynamics.
%%%%%%%%%%%%%%%%%%%%%%%%%%%%%%%%%%%%%%%%%%%%%%%%%%%%%%%
\begin{definition}[Generalized Dirac structure, \cite{Cordoni22}]\label{def:oc2} 
With the same notation as above, we call \textit{generalized stochastic Dirac
structure} a smooth vector subbundle $\mathcal{D}\subset T\mathcal{X}\oplus T^*\mathcal{X}$ such that $\mathcal{D}=\mathcal{D}^\perp$.
\end{definition}

%%%% These two definitions extend the foundational geometric principles of port-Hamiltonian systems to a stochastic setting, ensuring that their physical integrity, especially regarding \emph{energy flow}, is preserved even in the presence of uncertainty.

\begin{remark}
We note that the above Definition~\ref{def:oc2} of generalized stochastic Dirac structures via Stratonovich pairing is not canonically anchored in geometric mechanics. 
Using energy-preserving pathwise integrals to define orthogonality lacks local (fiber-wise) meaning and bypasses the axiomatic isotropy and maximality central to Dirac theory.
\\
To resolve this issue, one can replace the integral criterion with a structure-preserving pairing induced by a stochastic symplectic form on the It\^{o} tangent bundle. This pairing is defined as a bilinear form, denoted by
$\omega_{X}\bigl((\delta X,\sigma),(\delta Y,\rho)\bigr) := \langle \rho,\delta X \rangle - \langle \sigma,\delta Y \rangle$, and construct $D_x$ as the maximally isotropic subspace satisfying $\omega_X|_{D_x} \equiv 0$. This aligns with stochastic analogs of Courant algebroids.
\end{remark}

%%%%%%%%%%%%%%%%%%%%%%%%%%%%%%%%%%%%%%%%%%%
We define an \textit{implicit generalized stochastic port-Hamiltonian system} (IGSPHS), which naturally arises from extending Dirac structures to the stochastic setting. These generalized structures provide a geometric framework for capturing energy-conserving relations under stochastic perturbations. 
Classical % port-Hamiltonian systems 
PHSs use differential equations constrained by Dirac structures and driven by Hamiltonian functions representing stored energy. 
To model stochastic effects like noise or uncertainty, we introduce semimartingale perturbations and apply Stratonovich calculus, which aligns with the system's geometric nature. 
The IGSPHS ensures that the pair $\big(\delta X_t, \textbf{d}H(X_t)\big)$ lies in the Dirac structure $\mathcal{D}(X_t)$ at each time, preserving energy consistency under randomness. Extending this formulation to include resistive and controlled ports allows for modeling dissipation and external interactions. 
Resistive elements are described by a subbundle $\mathcal{R}$, while controlled ports represent inputs and outputs.

\begin{definition}[Implicit generalized stochastic PHS, \cite{Cordoni22}]
    Let $H\colon\mathcal{X}\to\mathbb{R}$ be a Hamiltonian function, 
    $Z$ a semimartingale perturbing the system, then an \textit{implicit generalized stochastic port-Hamiltonian system}
    on $\mathcal{X}$ is a 4-tuple $(\mathcal{X},Z,\mathcal{D},H)$ such that
    \begin{equation}
        \bigl(\delta X_t,\textbf{d}H(X_t)\bigr)\in\mathcal{D}(X_t)\quad 
        \forall\,t\in I.
    \end{equation}
        Including a resistive element and an external element control, then an \textit{implicit generalized port-Hamiltonian system with resistive structure} $\mathcal{R}$ is a 5-tuple $(\mathcal{X},\textbf{Z},\mathcal{F},\mathcal{D}, H)$ such that
    \begin{equation*}
        (-\delta X_t,\textbf{d}H,\delta f_t^R,e_t^R,\delta f_t^C,e_t^C)\in\mathcal{D(X)_t}\quad
        \text{with}\quad (\delta f_t^R,e_t^R)\in\mathcal{R(X)}_t.
    \end{equation*}
\end{definition}
A significant advantage of the port-Hamiltonian framework is its ability to track energy flow within a dynamical system. When adapting the framework to a stochastic context, verifying that the fundamental energy-consistency property is maintained is crucial.
Proposition~\ref{prop31} establishes that \textit{Interconnected Generalized Stochastic Port-Hamiltonian Systems} (IGSPHS) uphold a form of energy balance in both the pathwise and expectation senses.

The classical criterion of nonpositive energy dissipation through resistive ports is too strict in the presence of stochasticity. 
A more relaxed condition based on mean power balance ensures that energy dissipation occurs on average. This modification creates a robust, applicable framework for modeling stochastic energy exchange within the port-Hamiltonian paradigm.

%% OLD Text
%A key strength of the port-Hamiltonian framework is its ability to track energy flow through a system. When extended to the stochastic setting, it is essential to verify that this fundamental energy-consistency property is preserved. Proposition~\ref{prop31} shows that IGSPHS maintain a form of energy balance, both pathwise and in expectation. The classical condition of nonpositive energy dissipation through resistive ports is too strict in the presence of randomness; however, a weaker mean power balance ensures that energy dissipation holds on average.  This relaxation allows for a physically meaningful and practically applicable formulation of stochastic energy exchange.

%%%%%%%%%%%%%%%%%%%%%%%%%%%%%%%%%%%% Proposition 3.1.
\begin{proposition}[Energy balance of IGSPHS, \cite{Cordoni22}]\label{prop31}
    Implicit port-Hamiltonian systems satisfy an  energy conservation property that is
    \begin{equation}\label{cons}
        H(X_t)-H(X_0)=\int_0^t\langle \textbf{d}H,\delta X_s\rangle,
    \end{equation}
       or, in short notation 
    \begin{equation}
        \delta H(X_t)=\langle \textbf{d}H,\delta X_t\rangle .
    \end{equation}
    The energy balance is
    \begin{equation}
        H(X_t)-H(X_0)=\int_0^t\langle e_s^R,\delta f_s^R\rangle +\int_0^t\langle e_s^C,\delta f_s^C\rangle \le\int_0^t\langle e_s^C,\delta f_s^C\rangle .
    \end{equation}
However, the condition $\int_0^t\langle e_s^R,\delta f_s^R\rangle \le0$ imposed on the resistive port is too strong since it is difficult for it to happen in practice, so the idea is to introduce the following weaker definition of the resistive relation $\mathcal{R}_W\subset\mathcal{F}_{Z_R}\times\mathcal{E}_R$:
\begin{equation}
    \mathbb{E}\int_0^t\langle e_s^R,\delta f_s^R\rangle \le0.
\end{equation}
Consequently, the \textit{mean power balance} requires that the energy be conserved and dissipated in mean value, i.e.\ it reads
\begin{equation}
    \mathbb{E}\bigl(H(X_t)-H(X_0)\bigr) = \mathbb{E}\int_0^t\langle e_s^R,\delta f_s^R\rangle 
    +\mathbb{E}\int_0^t\langle e_s^C,\delta f_s^C\rangle \le\mathbb{E}\int_0^t\langle e_s^C,\delta f_s^C\rangle .
\end{equation}
\end{proposition}
%%%%%%%%%%%%%%%%%%%%%%%%%%%%%%%% end Proposition 3.1.

\begin{figure}[htbp]
\centering
\setlength{\unitlength}{0.6cm}
\includegraphics[width=.90\textwidth]{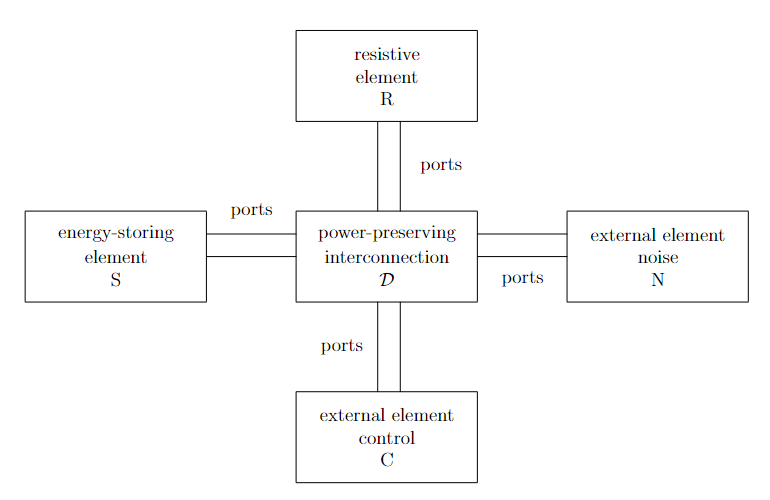}
\caption{Schematic representation of a general implicit port-Hamiltonian system.}\label{figure10}
\end{figure}

It is possible to further generalize the Hamiltonian by introducing an external perturbation of the system,
i.e.\ a new type of port called \textit{noise port} perturbed by the semimartingale $Z_N$ (see Figure~\ref{figure10}). 
Thus, in this case, the 
% \textit{implicit generalized stochastic port-Hamiltonian system with resistive structure} 
\textit{IGSPHS with resistive structure} 
is a 5-tuple $(\mathcal{X},\textbf{Z},\mathcal{F},\mathcal{D},H)$ such that
    \begin{equation}\label{64}
        \bigl(-\delta X_t,\textbf{d}H,\delta f_t^R,e_t^R,\delta f_t^C,e_t^C,\delta f_t^N,e_t^N\bigr)\in\mathcal{D}(X_t),
    \end{equation}
    and the weak energy balance is given by 
    \begin{equation}
        \mathbb{E}H(X_t) - \mathbb{E}H(X_0) \le \mathbb{E}\int_0^t\langle e_s^N,\delta f_s^N\rangle +\mathbb{E}\int_0^t\langle e_s^C,\delta f_s^C\rangle.
    \end{equation}

%%%%%%%%%%%%%%%%%%%%%%%%%%%%%%%%%%%

Furthermore, we can reformulate the SPHS in It\^{o} form by defining the following alternative representation of the Dirac structure.
%%%%%%%%%%%%%%%%%%%%%%%%%%%%%%
\begin{definition}[Dirac structure, \cite{Cordoni22}]
    Let $\mathcal{F}:=\mathcal{F}_{Z_R}\times\mathcal{F}_{Z_C}\times\mathcal{F}_{Z_N}$ be the space of
    flows $\delta f$, $\mathcal{E}=\mathcal{F}^*$ be the dual space of efforts $e$, $G_\theta\colon\mathcal{F}_{Z_\theta}\to\mathcal{F}_{Z_\theta}$ be a function such that $\langle e_t^S,G_\theta\delta f_t^\theta\rangle =\langle G_\theta^*e_t^S,\delta f_t^\theta\rangle$ and $J$ be a matrix with $J=-J^\top$. 
    Then the Dirac structure $\mathcal{D}$ can be defined as
    \begin{equation}
    \begin{split}
        \mathcal{D} := \bigl\{(&\delta f_t^S,\delta f_t^R,\delta f_t^C,\delta f_t^N,e_t^S,e_t^R,e_t^C,e_t^N)\in\mathcal{F}\times\mathcal{E}:\\
        &\delta f_t^S = -Je_t^S\delta Z_t-G_R\delta f_t^R-G_C\delta f_t^C-G_N\delta f_t^N,\\
        &e_t^R=G_R^*e_t^S,\quad 
        e_t^C=G^*_Ce_t^S,\quad 
        e_t^N=G^*e_t^S\bigr\}.
    \end{split}
    \end{equation}
\end{definition}
\begin{example}
Consider the special case 
\begin{equation}
\begin{split}
    &\delta f_t^R = -\Tilde{R}e_t^R\delta Z_t,\quad 
    \delta f_t^N = \xi_t\delta Z_t^N,\quad 
    \delta f_t^C = u_t\delta Z_t^C\\ &\text{with} \quad 
    \mathbb{E}\int_0^t\langle e_s^R,\Tilde{R}e_s^R\delta Z_s\rangle -\mathbb{E}\int_0^t\langle e_s^N,f_s^N\delta Z_s^N\rangle \ge0,
\end{split}
\end{equation}
where the reason for the minus sign in front of $\Tilde{R}$ is that we want it to be the incoming power regarding the interconnection (as in \cite{vanderSchaft14}).
\end{example}

%%%%%%%%%%%%%%%%%%%%%%%%%%%%%%
\begin{definition}[Stochastic input-output PHS with stochastic Dirac structure, \cite{Cordoni22}]
Using the same notation as above, if $\textbf{Z}=(Z,Z^R,Z^C,Z^N)$ is a semimartingale and $H\colon\mathcal{X}\to\mathbb{R}$ is a Hamiltonian function, then the \textit{stochastic input-output port Hamiltonian system with stochastic Dirac structure} is given by
    \begin{equation}\label{mio}
        \begin{cases}
            \delta X_t &= -JdH(X_t) \,\delta Z_t + G_R\Tilde{R}e_t^R \,\delta Z_t - G_Cu_t \,\delta Z_t^C - G_N\xi_t \,\delta Z_t^N,\\
            e_t^N &= G_N^*\textbf{d}H(X_t),\\
            e_t^C &= G_C^*\textbf{d}H(X_t)
        \end{cases}
    \end{equation}
    and by taking $\Tilde{J}=-J$ and $e_t^R=G^*_R\mathbf{d}H(X_t)$, the system \eqref{mio} becomes
     \begin{equation}\label{nonmio}
        \begin{cases}
            \delta X_t &= \big(\Tilde{J}+G_R\Tilde{R}G^*_R\big)\textbf{d}H(X_t) \,\delta Z_t - G_Cu_t\,\delta Z_t^C - G_N\xi_t \,\delta Z_t^N,\\
            e_t^N &= G_N^*\textbf{d}H(X_t),\\
            e_t^C &= G_C^*\textbf{d}H(X_t).
        \end{cases}
    \end{equation}
\end{definition}

%In stochastic modeling, the Stratonovich integral is often preferred because it is compatible with geometric structures, such as those found in PHS.
%However, it is usually more practical to work within the It\^{o} framework for analysis, simulation, and expectation-based reasoning purposes.
% The following 
Theorem~\ref{conv} provides an explicit conversion from the Stratonovich formulation of an IGSPHS
% implicit generalized stochastic port-Hamiltonian system 
to its equivalent It\^{o} representation.
% This transformation uses Lie derivatives to account for the correction terms introduced during conversion.
Under mild assumptions about the independence of the semimartingales driving the system, the resulting It\^{o} formulation separates the contributions of the stochastic, resistive, and controlled components.
% This enables a more precise analysis of system behavior and facilitates numerical integration and control design.

%%%%%%%%%%%%%%%%%%%%%%%%%%%%%%%%%%%%%%%%%%% Theorem 3.1
\begin{theorem}[Conversion of IGSPHS, \cite{Cordoni22}]\label{conv}
    If $X$ is a solution of the equation \eqref{nonmio} and $Z,Z^N,Z^C$ are such that 
    \begin{equation*}
        \langle Z,Z^C\rangle _t = \langle Z,Z^N\rangle _t = \langle Z,Z^C\rangle _t=0,
    \end{equation*}
    where $\langle \cdot,\cdot\rangle_t$ is the quadratic covariation at time $t$, 
    then $X$ can be equivalently rewritten in It\^{o} terms as
    \begin{equation}
        \begin{split}
            dX_t &= V^S(X_t)\,dZ_t + \mathcal{L}_{V^S}V^S(X_t) \,d\langle Z,Z\rangle_t\\
            &\qquad -\sum_{i=1}^{n^N} V_i^N(X_t) \,dZ_t^N
            - \frac{1}{2}\sum_{i,j=1}^{n^N}\mathcal{L}_{C_j^N} V_i^N(X_t) \,d\langle Z^{N;i},Z^{N;j}_t\rangle_t \\
            &\qquad -\sum_{i=1}^{n^C}V_i^C(X_t) u_t^i\,dZ_t^{C;i} 
            -\frac{1}{2}\sum_{i,j=1}^{n^C}\mathcal{L}_{V_j^C}V_i^C(X_t) u_t \,d\langle Z^{C;i},Z^{C;j}\rangle_t,
        \end{split}
    \end{equation}
    where $\mathcal{L}$ is the Lie derivative and $V^\alpha$, $\alpha=S,N,C$ are defined as
    \begin{equation}\label{V}
    \big(\Tilde{J}+G_R\Tilde{R}G^*_R\big)\,\textbf{d}H=V^S,\quad
    G_N\xi_t=\sum_{i=1}^{n^N}V_i^M,\quad
    G_C=\sum_{i=1}^{n^C}V^C_i.
\end{equation}
\end{theorem}

%%%%%%%%%%%%%%%%%%%%%%%%%%%%%%%%%%%%%%%%%%%%%%%%%%% Section 3
\section{Passivity in Stochastic Systems}\label{sec3}
Extending passivity to stochastic systems is challenging because noise directly influences the energy dynamics of the system.
In particular, the standard condition that the structure matrix $R$ be symmetric and positive semidefinite is insufficient in the stochastic setting.
The presence of a semimartingale $Z$ can introduce energy influx, rendering the system non-dissipative. Moreover, we need additional constraints on the stochastic perturbations to preserve losslessness and passivity.

To analyze the behavior of observables in I-S-O stochastic PHS,
% stochastic port-Hamiltonian systems with inputs and outputs (I-S-O),
it is essential to understand how smooth functions $\varphi \in C^\infty(\mathcal{X})$ evolve along the system trajectories. 
Proposition~\ref{prop41} shows that the evolution of such observables follows the same Lie bracket structure that defines the system dynamics. In this way, the geometric consistency of the Hamiltonian formulation is preserved under stochastic perturbations.
%, enabling energy-based analysis and control in uncertain environments.

%%%%%%%%%%%%%%%%%%%%%%%%%%%%%%%%%%%%%%%%%%%%%%%%%%%%%%% Proposition 4.1.
\begin{proposition}[Evolution along system trajectories, \cite{Cordoni22}]\label{prop41}
    If $X$ is the solution of an explicit I-S-O stochastic PHS with dissipation, i.e.
    \begin{equation}\label{eisosd}
        \begin{cases}
        \delta X_t &= X^L_H(X_t)\,\delta Z_t + uX^L_{H_g}(X_t)\,\delta Z^g_t + X^L_{H_N}(X_t)\,\delta Z^N_t,\\
        y_t &= [H,H_g]_L,
    \end{cases}
    \end{equation}
    with 
    \begin{equation*}
      X^L_H(\cdot) := [\cdot, H]_L,\quad 
      X^L_{H_g}(\cdot) := [\cdot, H_g]_L,\quad
      X^L_{H_N}(\cdot) := [\cdot, H_N ]_L,
    \end{equation*}
    then for all $\varphi\in C^\infty(\mathcal{X})$ it holds
    \begin{equation}\label{35}
        \begin{cases}
            \delta\varphi(X_t)&=[\varphi,H]_L(X_t)\,\delta Z_t + u[\varphi,H_g]_L(X_t)\,\delta Z_t^g+[\varphi,H_N]_L(X_t)\,\delta Z_t^N,\\
            y_t&=[H,H_g]_L.
        \end{cases}
    \end{equation}
\end{proposition}
%%%%%%%%%%%%%%%%%%%%%%%
% The equation changes when no external noise, and the semimartingale is perturbing the control, which is deterministic.
% The equation simplifies when there is no external noise and the control perturbation is deterministic.

To assess the stability and energy behavior of stochastic PHSs,
extending the classical concept of passivity to the stochastic setting is essential.
 %Due to randomness, system responses may fluctuate, so 
 We distinguish between two levels of passivity: strong passivity, which ensures energy bounds almost surely, and weak passivity, which guarantees these bounds in expectation, see Definition~\ref{def:passivity}.
% Definition~\ref{def:passivity} formalizes this distinction.

%%%%%%%%%%%%%%%%%%%%%
\begin{definition}[Strong and weak passivity, \cite{Cordoni22}]\label{def:passivity}
    If $H\in C^\infty(\mathcal{X})$ is the total energy of the explicit I-S-O stochastic PHS with dissipation,
    then it is \textit{strongly passive} if for all $t\ge0$ holds
    \begin{equation}\label{sp}
        H(X_t)\le H(X_0) + \int_0^t u^\top(s) y(s) \,\delta Z_s^C,
    \end{equation}
    or \textit{weakly passive} if for all $t\ge0$ it holds
    \begin{equation}\label{wp}
        \mathbb{E}H(X_t)\le\mathbb{E}H(X_0) + \mathbb{E}\int_0^t u^\top(s) y(s) \,\delta Z_s^C.
    \end{equation}
\end{definition}

%%%%%%%%%%%%%%%%%%%%%%%%%%%%%%%%%%%%%%
Now, assuming $\xi=0$, the energy conservation relation of the system 
\begin{equation}\label{81}
        \begin{cases}
            \delta X_t &= \big(J(X_t)-R(X_t)\big)\partial_xH(X_t)\,\delta Z_t + g(X_t)u\,\delta Z_t^C + \xi(X_t)\,\delta Z_t^N,\\
            e^C &= g^\top(X_t) \partial_x^H(X_t),
        \end{cases}
    \end{equation}
where $J=-J^\top$, is given by
\begin{equation}\label{calc}
    \begin{split}
        H(X_t)-H(X_0) &= \int_0^t\langle \textbf{d}H,\delta X_s\rangle
        = \int_0^t \langle e_s^R,\delta f_s^R\rangle 
           +\int_0^t \langle e_s^C,\delta f_s^C\rangle \\
        &= \int_0^t \bigl\langle \partial_xH(X_s),R(X_s)e^R\,\delta Z_s \bigr\rangle +\int_0^t \langle y,u\,\delta Z_s^C\rangle \\
        &= \int_0^t \bigl\langle \partial_xH(X_s),-R(X_s)\partial_xH(X_s)\delta Z_s\bigr\rangle 
        + \int_0^t y^\top u\,\delta Z_s^C\\
        &=-\int_0^t \partial_x^\top H(X_s)R(X_s)\partial_xH(X_s)\delta Z_s+\int_0^ty^\top u\,\delta Z_s^C.
    \end{split}
\end{equation}
%%%%%%%%%%%
Even if the matrix $R$ is strictly positive, this alone does not guarantee strong passivity.
In particular, we must also require the condition
\begin{equation}\label{adcond}
    \int_0^t \partial_x^\top H(X_s) R(X_s) \partial_x H(X_s) \,\delta Z_s \ge 0.
\end{equation}
However, satisfying this condition in practice is often unrealistic. 
A more tractable alternative is to impose the weaker requirement
\begin{equation}
    \mathbb{E}\int_0^t \partial_x^\top H(X_s) R(X_s) \partial_x H(X_s) \,\delta Z_s \ge 0,
\end{equation}
which ensures \textit{weak passivity}, cf.\ Def.~\ref{def:passivity}. 
However, working directly with expectations of Strato\-novich integrals can be technically challenging. 
To overcome this, it is natural to convert the expression in \eqref{wp} into It\^{o} form, by applying Theorem~\ref{conv} to equation \eqref{wp}, to exploit related probabilistic tools then.
%more accessible. The latter can be achieved by applying Theorem~\ref{conv} to equation \eqref{wp}.

%%%%%%%%%%%%%%%%%%%%%%%%%%%%%%%%%%%%%%%%%%%%%%%%%%%%%%%%%%%% Section 4
\section{Application and Generalization}\label{sec4}
\subsection{Interconnection of Multiple SPHS}\label{sec41}
An essential property of port-Hamiltonian systems (PHSs) is their interconnectivity, which allows complex systems to be viewed as compositions of simpler parts. 
The resulting interconnectivity regarding the components and how they are interconnected can be analyzed.
 In particular, through the composition of Dirac structures, the power-preserving interconnection of PHSs %port-Hamiltonian systems
 defines another PHS. 
The Hamiltonian of the interconnected PHS %port-Hamiltonian system
is the sum of the Hamiltonians of its components,  and the energy-dissipation relation is the union of the energy-dissipation relations of the subsystems.

The following Proposition~\ref{prop:51} discusses the connection of multiple SPHS, defining a new system with interconnected Dirac structures and combined Hamiltonians.

\begin{figure}[htbp]
\centering
\setlength{\unitlength}{0.8cm}
\includegraphics[width=.90\textwidth]{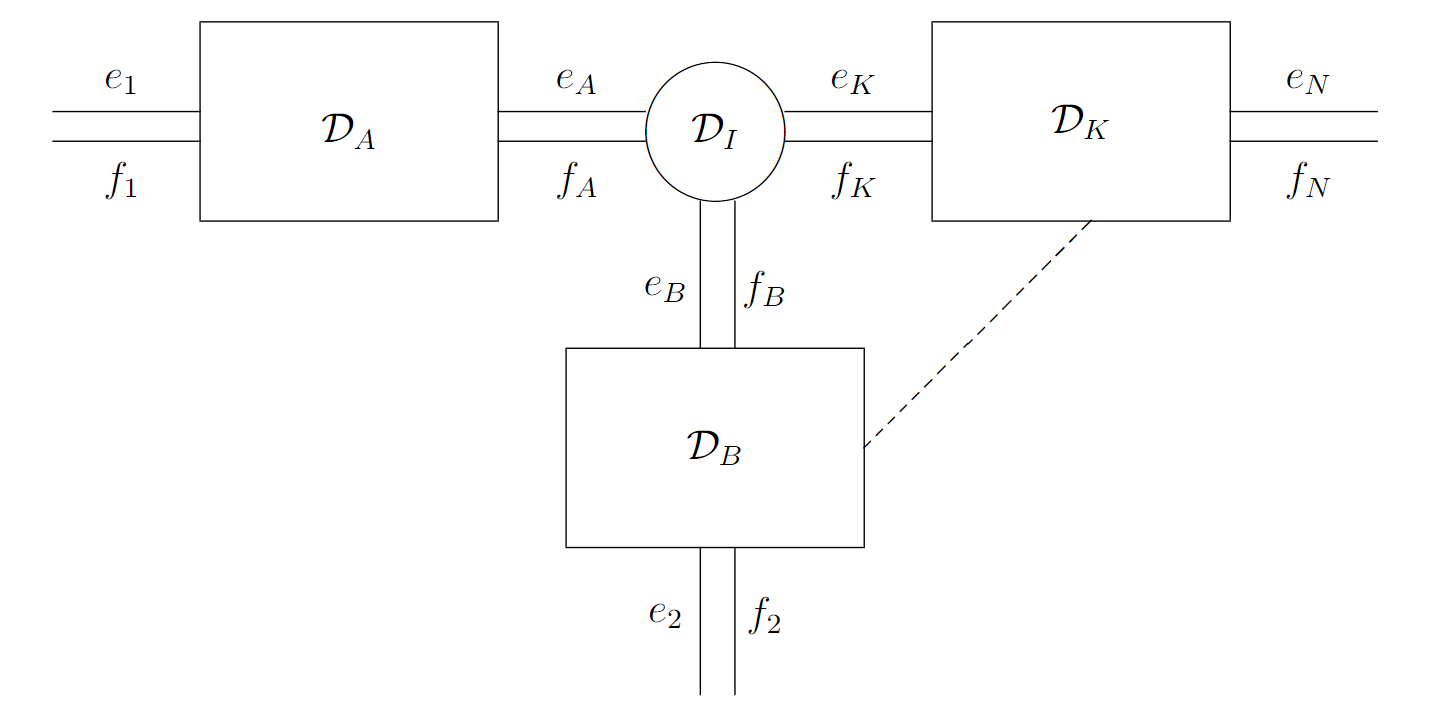}
\caption{Interconnection of $N$ implicit port-Hamiltonian systems.}\label{figure13}
\end{figure}

%%%%%%%%%%%%%%%%%%%%%%%%%%%%%%%%%%%%%%%%%%%% Proposition 5.1.
\begin{proposition}\label{prop:51}
Suppose we have $N$ stochastic port-Hamiltonian systems with state space $\mathcal{X}_i$, Hamiltonian $H_i$,
flow-effort space $\mathcal{F}_i\times\mathcal{E}_i$ and perturbation $\textbf{Z}^i$ for $i=1,\dots,N$. 
Assuming that they are connected by $\mathcal{D}_I$ (see Figure~\ref{figure13}), 
then their interconnection defines a stochastic port-Hamiltonian system with Dirac structure $\mathcal{D}\circ\mathcal{D}_I$ and Hamiltonian $H:=\sum_{i=1}^NH_i$.
\end{proposition}

%%%%%%%%%%%%%%%%%%%%%%%%%%%%%%%%%%%%%%%%

%%%%%%%%%%%%%%%%%%%%%%%%%%%%%%%%%%%%%%%%%%%%%%%%%%% Section 4.2.
\subsection{Discrete Stochastic PHS}\label{sec42}
Consider the continuous stochastic port-Hamiltonian system 
\begin{equation}\label{new}
    \begin{cases}
        dX_t &= \big((J-R)\partial_xH(X_t) + g(X_t)u_t\big)\,dt + \xi(X_t) \,\delta W_t,\\
        y_t &= g^\top(X_t)\partial_xH(X_t),\\
        z_t &= \xi^\top(X_t)\partial_xH(X_t),
    \end{cases}
\end{equation}
where $J =-J^\top$, R positive semidefinite, $g$ represents control port, $H$ is the Hamiltonian, $u\in U$ is the control
input, $y\in Y$ is the output of the system, $\xi$ is a matrix, $z$ is the associated effort to $\delta W_t$ and $W$ is a standard Brownian motion adapted to the reference filtration $(\mathcal{F}_t)_{t\ge0}$. Then we can introduce the discretization
\begin{equation}\label{def}
    \dot{X}(t_0^k + \tau h) = -f(t_0^k+\tau h) = -\sum_{j=1}^s f_j^k \,l_j(\tau),
\end{equation}
with
\begin{equation*}
    \dot{X}(t_i^k):=-f_i^k,\quad 
    l_i(\tau)=\prod_{j=1}^s\frac{\tau-c_j}{c_i-c_j},\quad 
    \tau\in[0,1],
\end{equation*}
where $l_i$ is the $i^{\rm th}$ Lagrange interpolation polynomial of order $s$ and $\tau$ is the normalized time parameterizing the sampling intervals. 
Thus, we can generalize the continuous SPHS to a discrete form, preserving the structure of the Hamiltonian and the control inputs, as follows:
%%%%%%%%%%%%%%%%%%%%%%
\begin{definition}[Discrete stochastic port-Hamiltonian system, \cite{Kotyczka19}]\label{def:dsphs}
    A \textit{discrete stochastic port-Hamiltonian system} can be written as
    \begin{equation}\label{dsphs}
        \begin{cases}
            X(t_0^k+c_ih) &= x_0^k - h\sum\limits_{i=1}^sa_{ij}f_j^k,\\
            X(t_0^k+h) &= x_0^k - h\sum\limits_{i=1}^sa_jf_i^k,\\
            -a_{ij}f^k &= (J_j^k-R_j^k)\,a_{ij}e^k_j+a_{ij}g_j^ku_j^k+b_{ij}\xi^k_j\,\Delta W,
        \end{cases}
    \end{equation}
    where $\Delta W$ is a truncated centered Gaussian random variable with variance $h$,
    $a_{ij}=\int_0^{c_i}l_j(\sigma)\,d\sigma$, 
    $a_j=\int_0^1l_j(\sigma)\,d\sigma$ and $M=M^\top$, cf.\  \cite{Kotyczka19}.
\end{definition}
Note that in the discrete case, the system is \textit{passive} if it holds
    \begin{equation}\label{eq:passive}
        \mathbb{E}\bigl[\Delta H^k\bigr]
        \le h\,\mathbb{E}\bigl[(y^k)^\top u^k\bigr].
    \end{equation}

%%%%%%%%%%%%%%%%%%%%%%%%%%%%%%%%%%%%%%%%%%%%%%%%%%%%%% Section 4.3.
\subsection{Stochastic Motion Model of Agents}\label{sec43}
Ehrhardt, Kruse, and Tordeux present an application to a stochastic motion model of agents \cite{Ehrhardt24}, who analyze the case where positions and velocities of agents are modeled in a ring structure. 
The initial positions and velocities are set, and the system is governed by differential equations involving the velocities and positions of neighboring agents \cite{Ehrhardt24}.
The dynamic equations of the agents are given by:
     \begin{equation}\label{motion}
    \begin{cases}
        dQ_n(t) &= \bigl(p_{n+1}(t)-p_n(t)\bigr)\,dt,\\
        dp_n(t) &= \bigl(U'(Q_n(t))-U'(Q_{n-q}(t))\bigr)\,dt \\ 
             &\qquad +\beta\bigl(p_{n+1}(t)-2p_n(t)+p_{n-1}(t)\bigr)\,dt + \sigma\,dW_n(t),
    \end{cases}
\end{equation}
with $Q(0)=Q_0\in[0,+\infty)^N$ the initial distance, $p(0)=p_0$ the initial velocity, $\beta\in(0,+\infty)$ a dissipation rate, $\sigma\in\mathbb{R}$ the noise volatility, 
$U'$ the derivative of a convex potential $U\in C^1(\mathbb{R},[0,+\infty))$ 
and $W=(W_n)_{n=1}^N\colon[0,+\infty)\times\Omega\to\mathbb{R}^N$ an $N$-dimensional standard Brownian motion defined on $(\Omega,\mathcal{F},\mathbb{P})$.
These equations represent the acceleration of the $n^{\rm th}$ agent depending on its neighbors' velocities and the Brownian motion's stochastic perturbations \cite{Gardiner85}.

The motion of the agents is further formulated using a stochastic port-Hamiltonian framework
\begin{equation*}
    dZ(t) = (J - R)\nabla H(Z(t))\,dt + G\,dW(t),
\end{equation*}
where $Z(t)=\big(Q(t),p(t)\big)^\top\in\mathbb{R}^{2N}$, $t\in[0,+\infty)$, $J$ and $R$ are defined as skew-sym\-me\-tric and symmetric positive semidefinite matrices, respectively. 
This formulation allows the application of Hamiltonian dynamics to model agents' behaviour under stochastic influences. 
In particular, the Hamiltonian is independent of $Q$ and $U$, and its expectation could increase with time. 
Moreover, describing the limiting behavior of these stochastic systems is challenging, so the authors \cite{Ehrhardt24}  focus on the specific scenario where the quadratic function characterizes the potential
\begin{equation}
    U(x)=\frac{(\alpha x)^2}{2}\quad x\in\mathbb{R},\; \alpha\in(0,\infty),
\end{equation}
in which the process reads 
\begin{equation}
    dZ(t)=BZ(t)\,dt + G\,dW(t),\quad Z(0)=(Q_0,p_0)^\top,
\end{equation}
where $B$ is defined such that $BZ(t)=(J-R)\nabla H(Z(t))$. 
In this case, the resulting process converges for $t\to\infty$ in distribution to a normal distribution with known expectation and covariance matrix.

%%%%%%%%%%%%%%%%%%%%%%%%%%%%%%%%%%%%% Section 6
%% \section{Results and Discussion}\label{sec5}
\section{Applications of Port-Hamiltonian Neural Networks}\label{sec5}
In this section, we examine the performance of \textit{port-Hamiltonian neural networks} (pHNNs) on a mass-spring system, both with and without damping.

Here, pHNNs provide a robust framework for learning the dynamics of non-auto\-nomous systems, which often involve time-dependent inputs and dissipative effects. These aspects typically pose challenges for conventional learning models. 

As Desai et al.\ \cite{Desai21} have shown, pHNNs can accurately model such systems and capture complex behaviors.
Notably, pHNNs can reconstruct Poincaré sections of chaotic systems, demonstrating their ability to learn the underlying structure from limited data, making pHNNs %particularly 
promising for applications involving nonlinear, forced, and damped systems, such as molecular dynamics, robotic control, and physical systems with unknown damping or input forces.

%% old text 
\begin{comment}
To introduce the port-Hamiltonian formalism into the neural network architecture, we can analyze the behavior of \textit{port-Hamiltonian Neural Networks} (pHNN) on different tasks such as the mass-spring system with the damped term, external force, and the Duffing equations. 
In particular, the % port-Hamiltonian Neural Network (pHNN)
pHNN is a significant advance in learning and predicting the dynamics of non-autonomous systems. 
Many real-world dynamical systems involve time-dependent forces and energy dissipation, challenging learning. 
% Desai, Roberts, Mattheakis, Sondak, and Protopapas \cite{Desai21} 
Desai et al.\ \cite{Desai21} evaluate the pHNN on several tasks, 
including a mass-spring system with damping and an external force, and a Duffing system. 
In particular, the pHNN can visually recover the Poincaré section of a chaotically driven system, 
highlighting its potential to identify and understand chaotic trajectories with minimal data. 
Its applicability extends to complex nonlinear forced and damped physical systems, 
encouraging applications in chemical bonding forces, robotic motion, and controlled dynamics without explicit knowledge of force and damping.
\end{comment}

\begin{figure}[htbp]
\centering
\includegraphics[width=.90\textwidth]{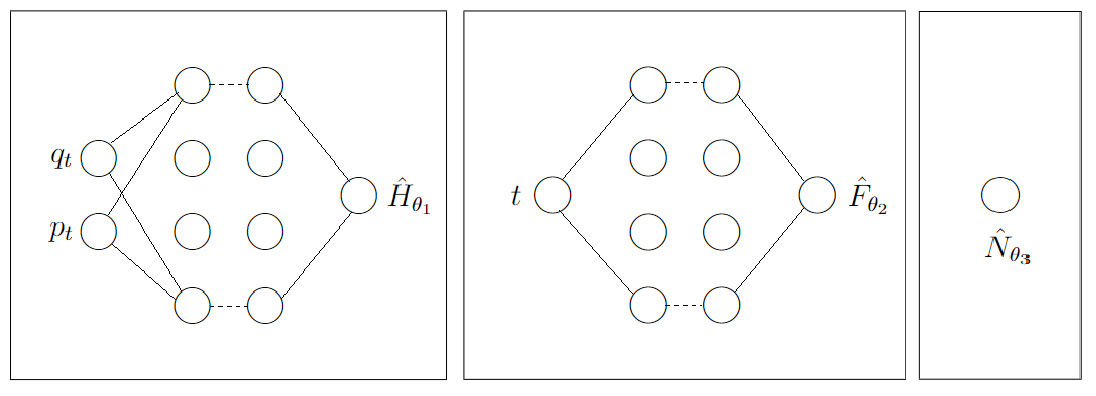}
\caption{Schematic representation of a port-Hamiltonian neural network (pHNN).}\label{figure4}
\end{figure}

As illustrated in Figure~\ref{figure4}, the core idea is to leverage port-Hamiltonian theory to explicitly learn the force term $F_{\theta_2}$,
the damping term $N_{\theta_3}$, and the Hamiltonian % function 
$\mathcal{H}_{\theta_1}$. 
This framework enables the prediction of the time derivatives:
\begin{equation}\label{mat}
    \begin{bmatrix}
        \hat{\dot{q}}_t\\
        \hat{\dot{p}}_t
    \end{bmatrix} 
    = \begin{bmatrix}
        \frac{d\hat{H}_{\theta_1}}{dp_t}\\
        -\frac{d\hat{H}_{\theta_1}}{dq_t}+\hat{N}_{\theta_3}\frac{d\hat{H}_{\theta_1}}{dp_t}+\hat{F}_{\theta_2}
    \end{bmatrix}.
\end{equation}

%%%%%%%%%%%%%%%%%%%%%%%%%%%%%%%%%%%
\subsection{First loss function}
We can define a loss function using the observed time derivatives $[\dot{q},\dot{p}]$ from the data, as described in \cite{Wang22}:
\begin{comment}
\begin{equation}\label{loss}
    \mathcal{L}=\underbrace{||\hat{\dot{q}}_t-\dot{q}_t||_2^2+||\hat{\dot{p}}_t-\dot{p}_t||_2^2}_{\text{first part}}
     + \underbrace{\lambda_F||\hat{F}_{\theta_2}||_1 + \lambda_N||\hat{N}_{\theta_3}||_1}_{\text{second part}},
\end{equation}
\end{comment}

\begin{equation}\label{loss}
    \mathcal{L}_{qp} = ||\hat{\dot{q}}_t-\dot{q}_t||_2^2
    + ||\hat{\dot{p}}_t-\dot{p}_t||_2^2.
\end{equation}

%%%%%%%%%%%%%%%%%%%%%%%%%%%%%%%%%%%%
\subsection{Second loss function}
Recall that the following It\^{o} SDE can model a stochastic port-Hamiltonian system
\begin{equation*}
   dX_t=f(X_t)\,dt+\Sigma(X_t)\,dW_t,\qquad X_t=[q_t^\top,p_t^\top]^\top,
\end{equation*}
with Brownian motion $W_t$. Over one sampling interval $\Delta t$ the Euler--Maruyama increment $\Delta X_t=X_{t+\Delta t}-X_t$ satisfies $E[\Delta X_t]=f(X_t)\Delta t+O(\Delta t^{2})$ and $\text{cov}[\Delta X_t] = \Sigma\Sigma^\top (X_t) \Delta t+O(\Delta t^{2})$. 
Substituting neural predictions $\hat f_\theta$ and $\hat\Sigma_\theta$ for the unknown drift and diffusion yields the loss
\begin{equation}\label{stoch_loss}
    \mathcal{L}_{\text{DDpHNN}}=\frac{1}{\Delta t}\lVert\Delta X_t-\hat f_\theta\Delta t\rVert_2^{2}+\frac{1}{\Delta t}\lVert\Delta X_t\Delta X_t^\top -\hat\Sigma_\theta\hat\Sigma_\theta^\top \Delta t\rVert_F^{2},
\end{equation}

The drift-diffusion loss function \eqref{stoch_loss} combines a drift term and a diffusion term: % ('drift-diffusion loss'): 
\begin{equation*}
\mathcal{L}_{\text{DDpHNN}} 
=\underbrace{\frac{1}{\Delta t}\|\Delta X_t-\hat f_\theta\Delta t\|_2^{2}}_{\mathcal{L}_{\text{drift}}} 
+\underbrace{\frac{1}{\Delta t}\|\Delta X_t\Delta X_t^{\!\top}-\hat\Sigma_\theta\hat\Sigma_\theta^{\!\top}\Delta t\|_F^{2}}_{\mathcal{L}_{\text{diff}}}. 
\end{equation*}

We scale the second term of the drift-diffusion loss by $\lambda$ to rebalance the training gradients and prevent the model from exaggerating the predicted noise magnitude, all while leaving drift learning unaffected. This results in well-calibrated diffusion without changing the network architecture
\begin{equation}
     \mathcal{L}_{\text{DDpHNN}}^{(\lambda)} =\mathcal{L}_{\text{drift}} +\lambda\,\mathcal{L}_{\text{diff}},
\end{equation}
and to prevent the unbounded growth of $\Sigma_\theta$, we add a $\ell_{1}$-term ($\alpha << 1$) to control the diffusion amplitudes
\begin{equation}\label{stoch_loss_ddphnn}
\mathcal{L}_{\text{DDpHNN}}^{(\lambda,\alpha)} 
=\mathcal{L}_{\text{drift}} 
+\lambda\,\mathcal{L}_{\text{diff}} 
+\alpha\lVert\hat\Sigma_\theta\rVert_{1}.   
\end{equation}

Unlike standard HNN/pHNN training, which only accounts for deterministic drift and usually disregards stochasticity, \eqref{stoch_loss_ddphnn} explicitly learns a state-dependent diffusion $\Sigma_\theta$ by matching the first two moments of Euler-Maruyama increments. 
This moment matching, together with the weighted and $\ell_{1}$-regularized variant $\mathcal{L}_{\text{DDpHNN}}^{(\lambda,\alpha)}$, promotes calibrated noise magnitudes while leaving drift learning intact, hence providing 
one of the novel contributions of this paper.  Typically, neural-SDE models learn drift and diffusion using either moment-matching of Euler-Maruyama increments or likelihood-based training. 
For example, \cite{Dietrich23} fit drift/diffusion via one-step EM maximum likelihood, while \cite{Vaisband25} proposes a hybrid objective that outperforms pure moment-matching. 
Our contribution is to introduce this approach into a port-Hamiltonian setting with a weighted moment-matching loss and an $\ell_1$ penalty on $\Sigma_\theta$, a method which, as far as we know, has not been reported before.

%%%%%%%%%%%%%%%%%%%%%%%%%%%%%%%%%%%%%
\subsection{Third loss function}
Let a minibatch $B=\{(x_i,f_i,h_i)\}_{i=1}^m$ be given, with the states $x_i=(q_i,p_i)$, ground truth vector field $f_i=(\dot q_i,\dot p_i)$, and ground truth energy $h_i=H_{\text{true}}(x_i)$.
In our model, we define $H_\theta(x)$ and the Hamiltonian vector field
\begin{equation*}
f_\theta(x)=J\nabla_x H_\theta(x)=\big(\tfrac{\partial H_\theta}{\partial p},\;-\tfrac{\partial H_\theta}{\partial q}\big),\quad
J=\begin{bmatrix}0&I\\-I&0\end{bmatrix}.
\end{equation*}

We build a new loss function from the following three terms:
%%%%%%%%%%%%%%%%%
\subsubsection*{Vector field MSE}
\begin{equation*}
\mathcal{L}_{\text{vf}}
=\frac{1}{m}\sum_{i=1}^m \big\| f_\theta(x_i)-f_i \big\|_2^2 .
\end{equation*}
this serves to predict the Hamiltonian vector field $(\dot q, \dot p) = \left(\frac{\partial H}{\partial p}, -\frac{\partial H}{\partial q}\right)$ and minimize MSE to the ground-truth $(\dot q, \dot p)$.
This is the core loss used by Hamiltonian Neural Networks (HNN) \cite{Greydanus2019HNN} and follow-ups that supervise time derivatives rather than states.

%%%%%%%%%%%%%%%%%%%%%%%%
\subsubsection*{Batch centered energy MSE}
We regress the scalar energy $H(x)$, but we compare the centered values $H-\bar H_{\text{batch}}$, so the loss is invariant to an additive constant. 
In Hamiltonian mechanics, the equations of motion depend on gradients of $H$, so adding a constant to $H$ changes nothing physically. 
Centering makes the regression immune to unknown offsets while teaching the network the shape and scale of the energy landscape.
\begin{equation*}
\bar H_\theta=\frac{1}{m}\sum_{i=1}^m H_\theta(x_i),\qquad
\bar h=\frac{1}{m}\sum_{i=1}^m h_i,
\end{equation*}
%%%%%%%%%%%%%%%%%%%%%%
\begin{equation*}
\mathcal{L}_{\text{E}}
=\frac{1}{m}\sum_{i=1}^m \Big( \big(H_\theta(x_i)-\bar H_\theta\big)-\big(h_i-\bar h\big) \Big)^2 .
\end{equation*}

%%%%%%%%%%%%%%%%%%%%%%%
\subsubsection*{One period symplectic rollout MSE}
For a set of $K$ one-period rollouts $\{X^{(r)}_t\}_{t=0}^L$ with step $\Delta t$ (so $T=L\Delta t$), generate a predicted path $\{\hat X^{(r)}_t\}$ with the implicit midpoint update
\begin{equation*}
\hat X^{(r)}_{t+1}=\hat X^{(r)}_{t}+\Delta t\; J\nabla H_\theta\!\left(\tfrac{\hat X^{(r)}_{t}+\hat X^{(r)}_{t+1}}{2}\right),
\quad \hat X^{(r)}_{0}=X^{(r)}_{0},
\end{equation*}
and define
\begin{equation*}
\mathcal{L}_{\text{roll}}
=\frac{1}{K}\sum_{r=1}^K \frac{1}{L+1}\sum_{t=0}^{L}
\big\| \hat X^{(r)}_{t}-X^{(r)}_{t} \big\|_2^2 .
\end{equation*}

$\mathcal{L}_{\text{roll}}$ serves for long-horizon trajectory fidelity. The total loss is:
\begin{equation}
\mathcal{L}_{VER}
= \mathcal{L}_{\text{vf}}
+ \lambda_{\text{E}}\,\mathcal{L}_{\text{E}}
+ \mathcal{L}_{\text{roll}},   
\end{equation}
with $\lambda_{\text{E}}>0$ a tuning weight.

The VER objective ($\mathcal{L}_{VER}$) unifies three complementary signals-vector-field MSE, batch-centered energy regression, and a symplectic rollout penalty, into a single loss.
To our knowledge this exact combination has not appeared before. This combination is interesting because it simultaneously constrains local derivatives, the global energy landscape, and long-horizon trajectory fidelity under a structure-preserving integrator. 
The VER loss combines three terms: a vector-field MSE that matches $(\dot q,\dot p)=J\nabla H_\theta$ to the observed derivatives, a batch-centered energy regression comparing $H_\theta-\bar H_\theta$ to $H_{\text{true}}-\bar h$ to make the target invariant to additive constants while shaping the energy landscape, and a one-period symplectic rollout penalty that enforces long-horizon fidelity. 
The first component follows the HNN paradigm of supervising time derivatives, as proposed by $H$ \cite{Greydanus2019HNN}, and the third uses "symplectic in the loss" training with structure-preserving schemes \cite{DavidMehats23}. 
To our knowledge, unifying all three in a single objective is novel, as it jointly constrains local dynamics, global energy structure, and multi-step stability.

We will now test the three objectives. The resulting learning curves, phase portraits, energy traces, state trajectories, and error statistics are presented in the following figures and discussed below.

%%%%%%%%%%%%%%%%%%%%%%%%%%%%%
\subsection{Tests and Results}
\subsubsection{Comparing $\mathcal{L}_{qp}$ and $\mathcal{L}_{DDpHNN}$:}
We test the first and second objectives on the canonical damped-mass-spring system.
\subsubsection*{Training loss}
\begin{figure}[H]
\centering
\includegraphics[width=0.52\textwidth]{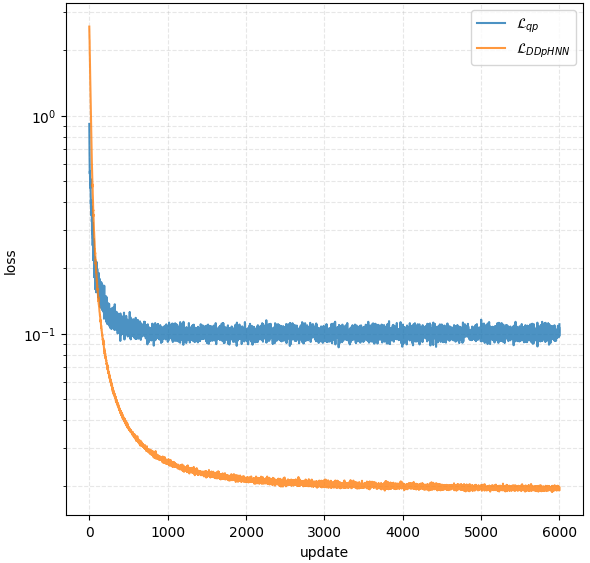}
\caption{Training-loss curves.}\label{det_vs_stoch_loss}
\end{figure}

Figure~\ref{det_vs_stoch_loss} confirms stable convergence for both objectives. 
The loss $\mathcal{L}_{\text{DDpHNN}}$ (orange) decays one order of magnitude below the deterministic loss $\mathcal{L}_{\text{qp}}$ (blue) and plateaus without signs of overfitting.
This shows that the additional diffusion term is well behaved.

%%%%%%%%%%%%%%%%%%%%%%%%%%%%%
\subsubsection*{Phase portrait and energy evolution}
\begin{figure}[H]
  \centering
  \begin{subfigure}[b]{0.49\textwidth}
    \centering
    \includegraphics[width=\linewidth]{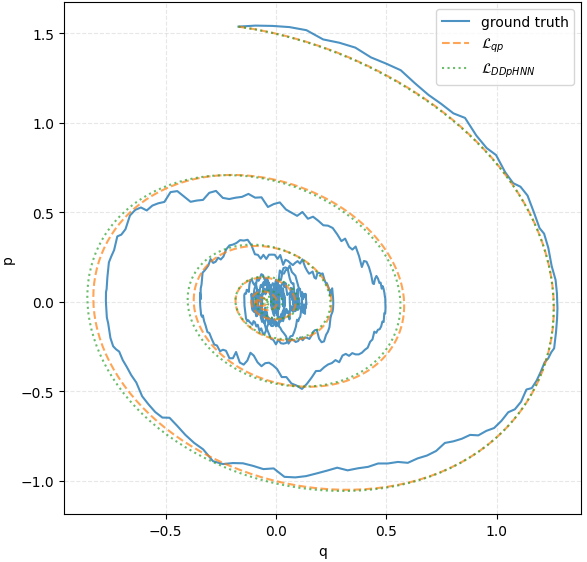}
    \caption{Phase portrait}
    \label{phase_space}
  \end{subfigure}
  \hfill
  \begin{subfigure}[b]{0.485\textwidth}
    \centering
    \includegraphics[width=\linewidth]{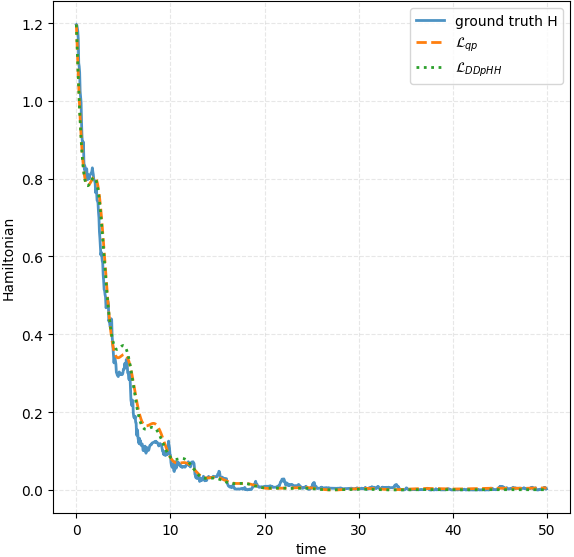}
    \caption{Energy evolution}
    \label{energy_evolution}
  \end{subfigure}
  \caption{Phase portrait and corresponding energy evolution of the undamped mass-spring system.}
  \label{fig:portrait_energy}
\end{figure}

As shown in Figure~\ref{phase_space}, both models closely follow the true spiral trajectory for the entire time window. 
Figure~\ref{energy_evolution} tracks the Hamiltonian $H(q,p)$. Both networks exhibit the dissipative trend.

%%%%%%%%%%%%%%%%%%%%%%%%%%%%%
\subsubsection*{Position and momentum}
\begin{figure}[H]
\centering
\includegraphics[width=1\textwidth]{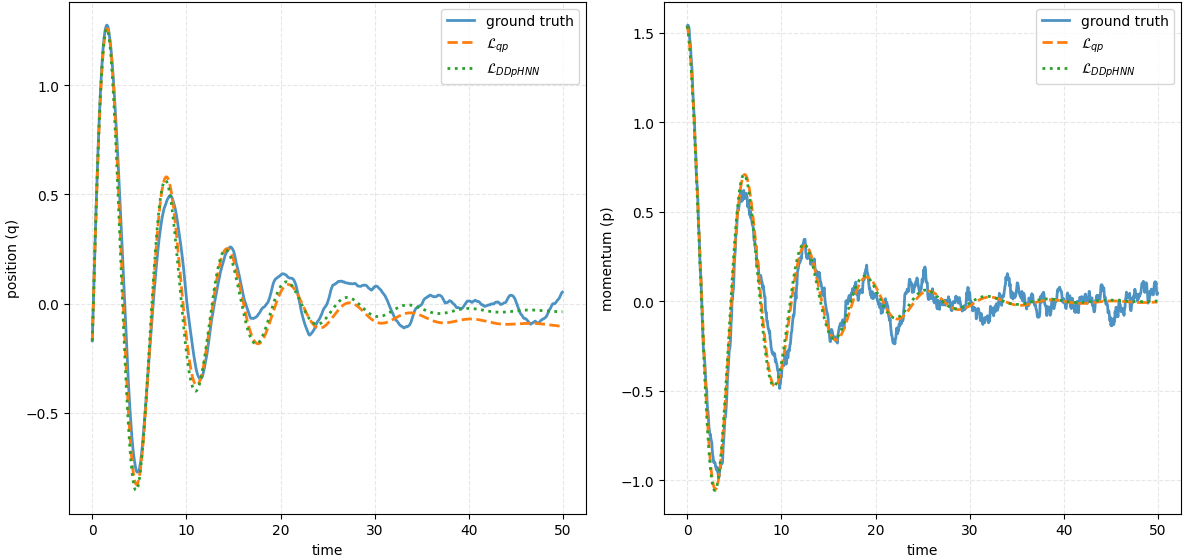}
\caption{Position and momentum.}\label{position_and_momentum}
\end{figure}
All three trajectories begin with the same damped oscillatory pattern and gradually diverge as the rollout progresses, demonstrating a consistent decrease in amplitude.

\begin{comment}
We now repeat the experiment on an undamped mass-spring system training the same architecture with both $\mathcal{L}_{qp}$ and $\mathcal{L}_{DDpHNN}$. The corresponding results are presented in the following Figure~\ref{fig:undamped_system}.
\begin{figure}[H]
    \centering
    \begin{subfigure}[b]{0.49\textwidth}
        \centering
        \includegraphics[width=\textwidth]{img/q_no_damping.png}
        \caption{Position}
        \label{fig:undamped_q}
    \end{subfigure}
    \hfill 
    \begin{subfigure}[b]{0.49\textwidth}
        \centering
        \includegraphics[width=\textwidth]{img/p_no_damping.png}
        \caption{Momentum}
        \label{fig:undamped_p}
    \end{subfigure}
    \caption{Evolution of an undamped mass-spring system.}
    \label{fig:undamped_system}
\end{figure}

\begin{remark}
Unlike structure-preserving \cite{Greydanus2019HNN} or physics‑informed \cite{Raissi2019PINN} formulations, which enforce energy conservation for an undamped mass-spring oscillator, these networks here only empirically learn the flow.  
Consequently, once they deviate from the  actual constant energy orbit, nothing forces them back. The predicted amplitudes in Figure~\ref{fig:undamped_system} gradually drift above (stochastic model) or below (deterministic model) the ground truth trajectory. A forthcoming paper \cite{Youness2} remedies this issue by embedding the port‑Hamiltonian constraints directly into a stochastic port Hamiltonian network. 
\end{remark}
\end{comment}

\subsubsection{Comparing $\mathcal{L}_{VER}$ to a baseline Multi-Layer Perceptron (MLP)}
\subsubsection*{The 3-DOF robotic arm}
We consider a 3-DOF arm (three joints). For brevity, we present Joint 1 plots, results for Joints 2-3 are similar.
\begin{figure}[H]
\centering
\includegraphics[width=0.8\textwidth]{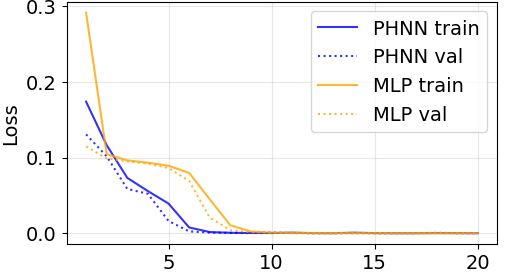}
\caption{Training-loss curves PHNN vs baseline model (MLP).}\label{}
\end{figure}

PHNN converges faster and lower than the MLP, training and validation track closely which indicate no overfitting.

\begin{figure}[H]
\centering
% Row 1: top-left (p1), top-right (MAE)
\begin{subfigure}[t]{0.49\textwidth}
\centering
\includegraphics[width=\linewidth]{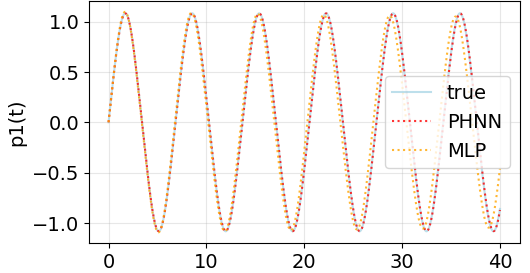}
\caption{$p_1(t)$}
\end{subfigure}
\hfill
\begin{subfigure}[t]{0.49\textwidth}
\centering
\includegraphics[width=\linewidth]{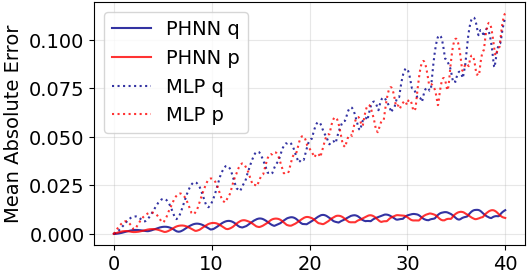}
\caption{Mean absolute error}
\end{subfigure}

\vspace{0.4em}

% Row 2: bottom-left (q1), bottom-right (phase portrait)
\begin{subfigure}[t]{0.49\textwidth}
\centering
\includegraphics[width=\linewidth]{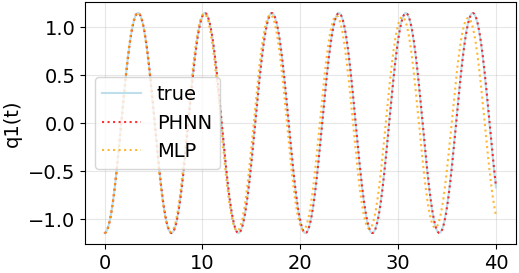}
\caption{$q_1(t)$}
\end{subfigure}
\hfill
\begin{subfigure}[t]{0.49\textwidth}
\centering
\includegraphics[width=\linewidth]{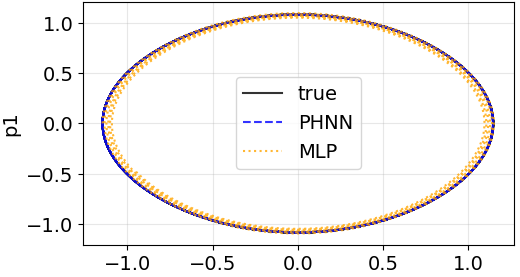}
\caption{Phase portrait $(q_1,p_1)$}
\end{subfigure}

\caption{PHNN vs MLP: time series, error, and phase portrait.}
\label{fig:three-dof-grid}
\end{figure}

The PHNN trajectories are nearly indistinguishable from the ground truth. However, the MLP shows phase lag. The PHNN tracks amplitude and phase closely, while the MLP accumulates phase error. PHNN errors remain low and consistent, whereas MLP errors increase over time.

And now we do the same tests but for a longer time interval $[0,120]$:

\begin{figure}[H]
\centering

% Row 1: top-left (p1), top-right (MAE)
\begin{subfigure}[t]{0.49\textwidth}
\centering
\includegraphics[width=\linewidth]{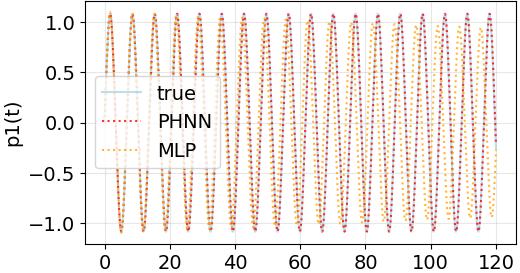}
\caption{$p_1(t)$}
\end{subfigure}
\hfill
\begin{subfigure}[t]{0.49\textwidth}
\centering
\includegraphics[width=\linewidth]{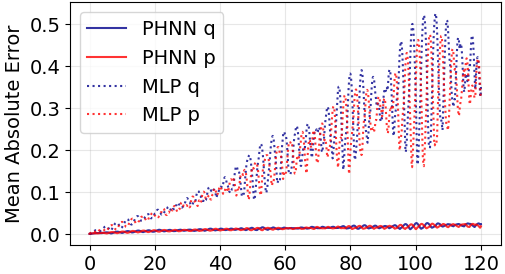}
\caption{Mean absolute error}
\end{subfigure}

\vspace{0.4em}

% Row 2: bottom-left (q1), bottom-right (phase portrait)
\begin{subfigure}[t]{0.49\textwidth}
\centering
\includegraphics[width=\linewidth]{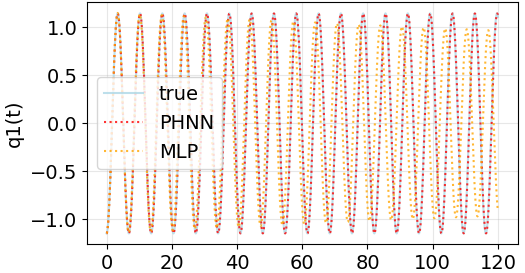}
\caption{$q_1(t)$}
\end{subfigure}
\hfill
\begin{subfigure}[t]{0.49\textwidth}
\centering
\includegraphics[width=\linewidth]{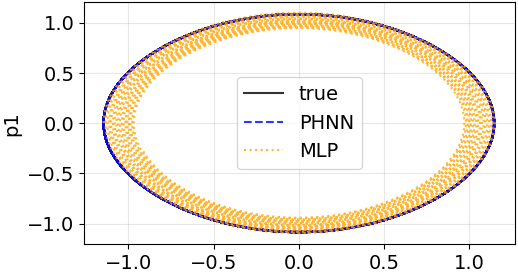}
\caption{Phase portrait $(q_1,p_1)$}
\end{subfigure}

\caption{PHNN vs MLP (extended): time series, error, and phase portrait.}
\label{fig:three-dof-grid}
\end{figure}

%%%%%%%%%%%%%%%%%%%%%%%%%%%%%%%%%%%%%%%%%%%%
\subsubsection*{The Duffing Oscillator}

\begin{figure}[H]
\centering
\includegraphics[width=0.8\textwidth]{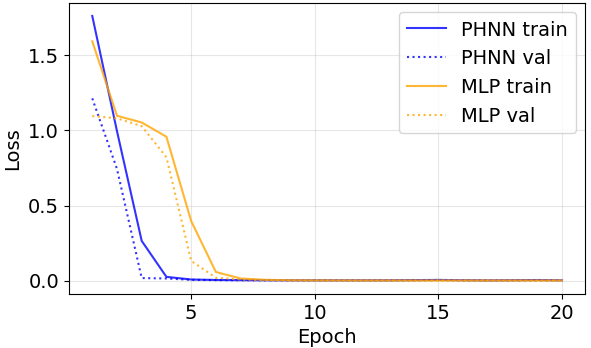}
\caption{Training-loss curves PHNN vs baseline model (MLP).}
\end{figure}

Both models drop quickly. PHNN converges faster and to a lower level. The training and validation curves track each other, indicating that there is no overfitting. 

\begin{figure}[H]
\centering
\includegraphics[width=0.8\textwidth]{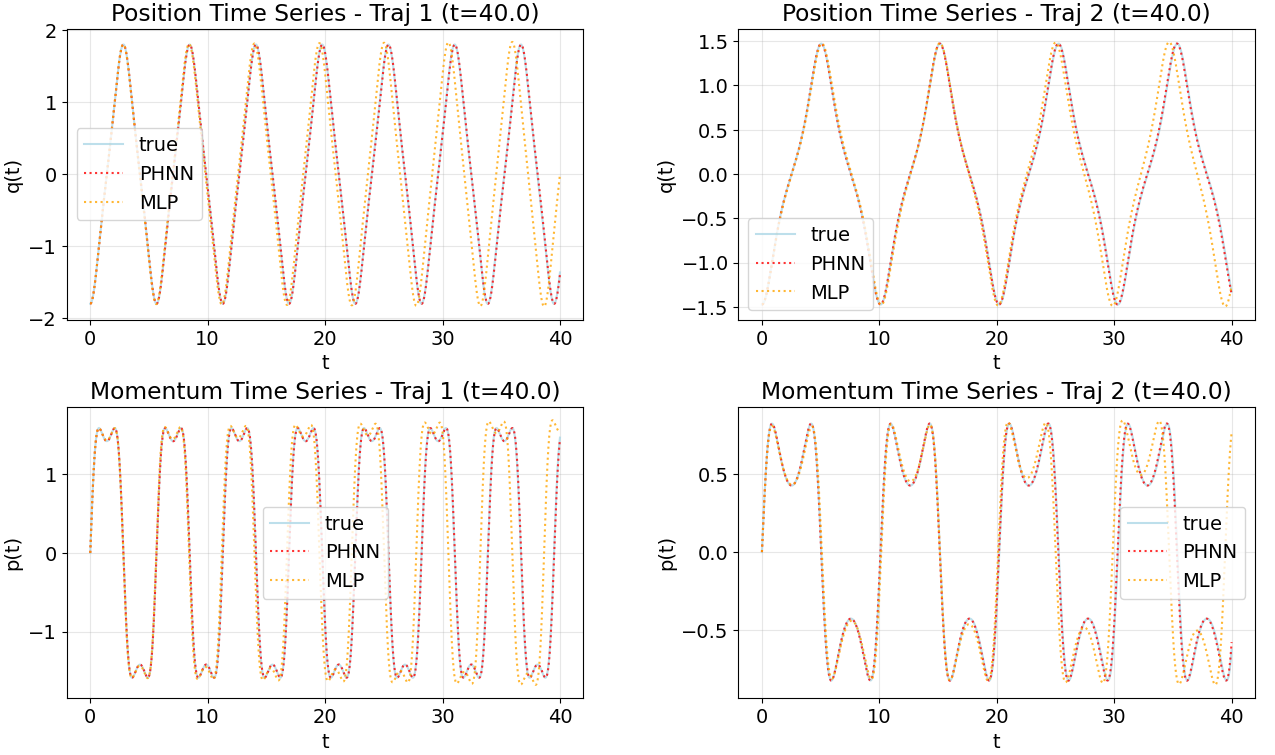}
\caption{Evolution of $q$ and $p$ for two different trajectories.}
\end{figure}

\begin{figure}[H]
\centering
\includegraphics[width=1\textwidth]{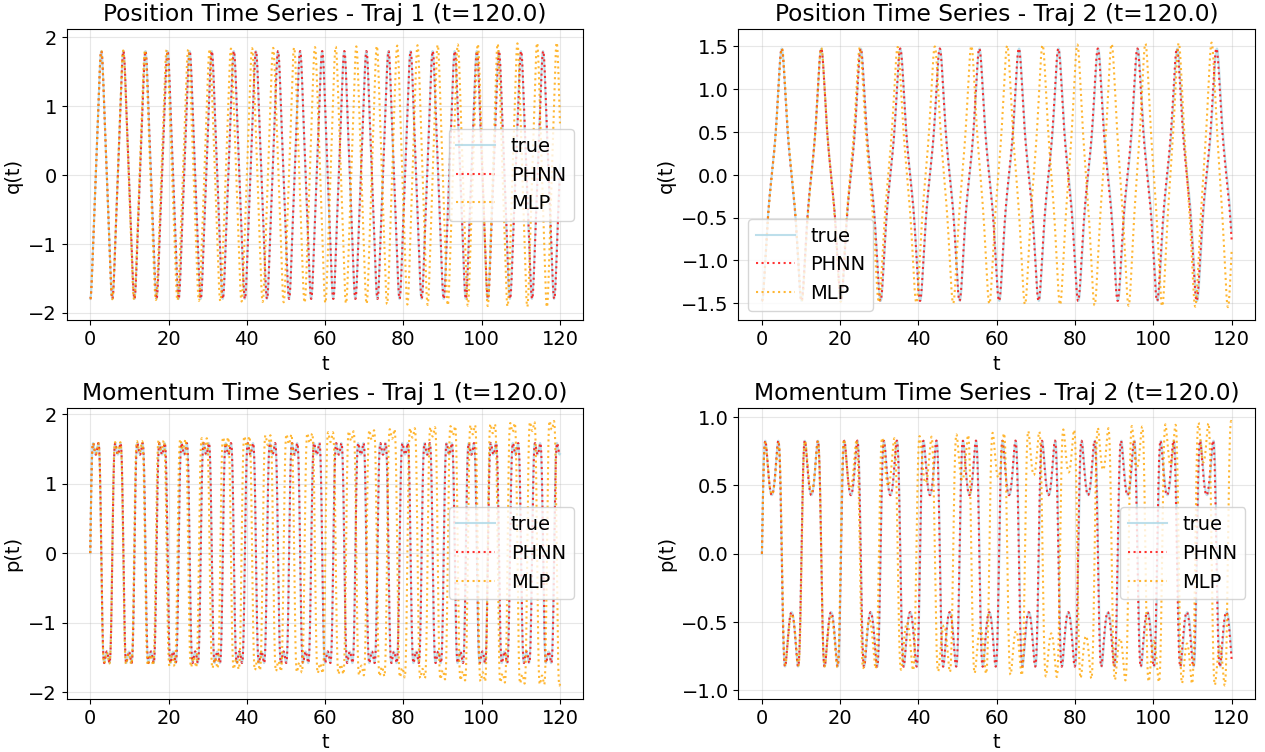}
\caption{Evolution of $q$ and $p$ for two different trajectories in a longer time interval $[0,120]$.}
\end{figure}

\begin{figure}[H]
\centering
\includegraphics[width=1\textwidth]{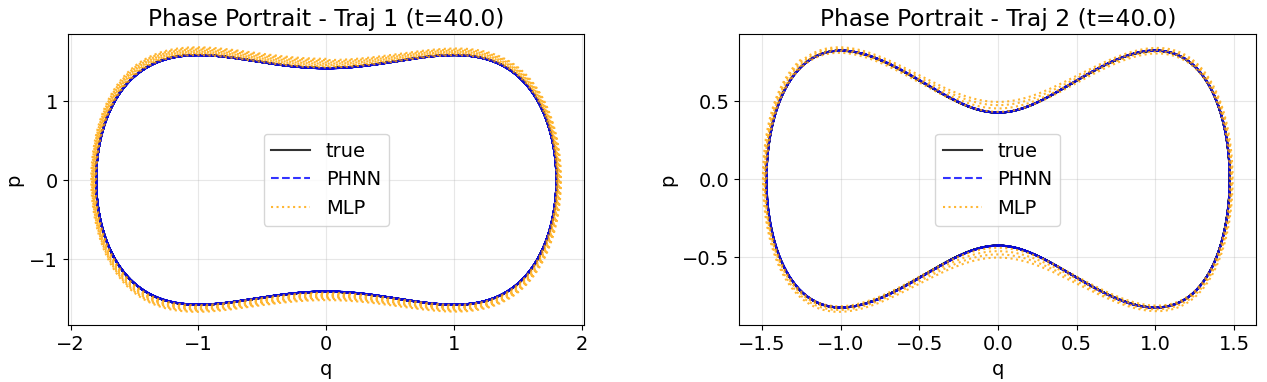}
\caption{Phase portrait for two different trajectories.}
\end{figure}

The PHNN is on top of the closed ground truth orbits. The MLP is close, but shows minor phase distortions near the turning points, hence demonstrating that both models have good short-horizon fidelity. 

\begin{figure}[H]
\centering
\includegraphics[width=1\textwidth]{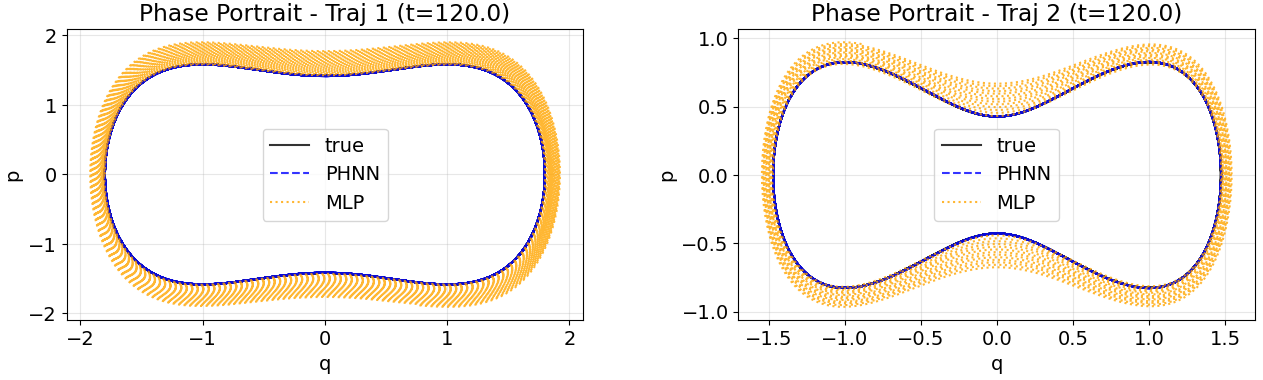}
\caption{Phase portrait for two different trajectories and a longer time interval $[0,120]$}
\end{figure}

The PHNN aligns well. While the PHNN stays synchronized over the extended horizon, the MLP accumulates phase lag and amplitude drift. 

\begin{figure}[H]
\centering
\includegraphics[width=1\textwidth]{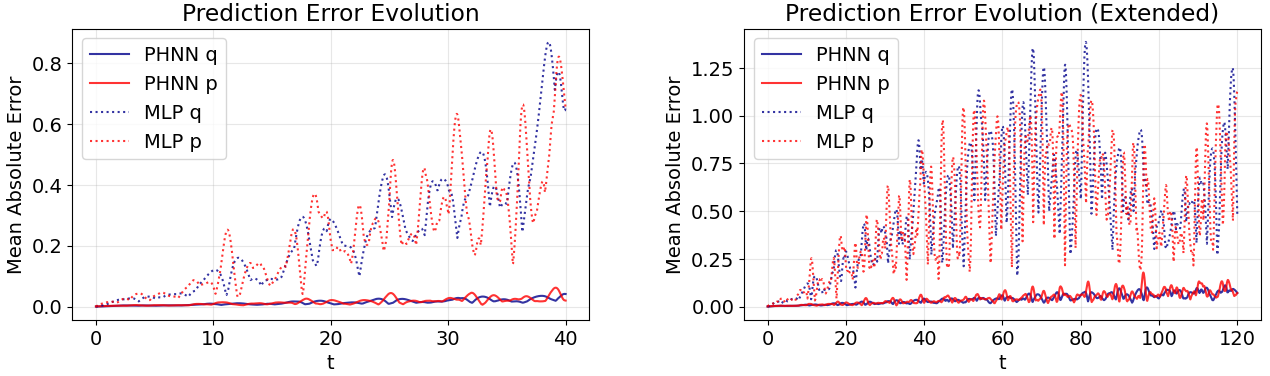}
\caption{Mean Absolute Error.}
\end{figure}

PHNN's $q$ and $p$ errors remain low and bounded over time, MLP's errors grow progressively, reflecting dephasing miscalibration during long rollouts.

\section{Conclusions and Outlook}\label{sec6}
The method of Colonius and Gr\"une \cite{Colonius02} 
uses a neural network to estimate a gradient field for controller design, removing the necessity to solve complicated partial differential equations. 
To extend this framework to stochastic port-Hamiltonian systems (SPHS), however, we must incorporate stochastic elements into the system model and the control synthesis.

First, we defined a stochastic version of the port-Hamiltonian system using stochastic differential equations (SDEs), such as:
\begin{equation*}
dx = \bigl(J(x) - R(x)\bigr)\, \nabla H(x)\, dt 
    + G(x) u \,dt + \Sigma(x)\, dW, 
\end{equation*}
Here, $\Sigma(x)$ is a matrix that captures the noise intensity, and $dW$ represents a Wiener process.
Second, the control strategy must be adapted to handle stochasticity. 
A modified control law can be written as follows
 \begin{equation*}
  u = -K(x) + \mu(x)
\end{equation*}
Here, $K(x)$ is derived from the deterministic system, and $\mu(x)$ compensates for stochastic effects.
A suitable Lyapunov function is constructed and analyzed using It\^{o}'s calculus to ensure system stability.
The objective is to demonstrate that the expected value of the Lyapunov function decreases over time, 
guaranteeing stability under stochastic dynamics.
Evolutionary optimization strategies should be adapted for this stochastic setting by redefining the fitness function to reflect expected performance across a range of noise conditions.
The ball-and-beam system will be extended to include stochastic disturbances as a practical demonstration. 
Simulations will illustrate the effects of noise and validate the proposed stochastic control approach.

%%%%%%% Outlook  %%%%%%%
An important direction for future research is to extend our conversion framework beyond the assumption of pairwise uncorrelated stochastic drivers, as required by Theorem~\ref{conv}. 
This assumption excludes a wide class of practical systems with inherently coupled control and noise inputs. 
Future work could address this limitation and enable a more general and realistic modeling of coupled stochastic systems
by incorporating non-orthogonal stochastic drivers using the full cross-variation tensor, $\langle Z^i, Z^j \rangle$ . 
This approach would require applying It\^{o}'s Lemma in manifold coordinates equipped with a nontrivial Levi-Civita connection. %This would enable a more general and realistic modeling of coupled stochastic systems.

\appendix
%%%%%%%%%%%%%%%%%%%%%%%%%%%%%%%%%%%%%%%%%%%%%%%%%%%%%%%
\section{Stochastic Neural Networks}\label{appA}
Neurons are the basic computational units of the brain and form neural networks (NNs) through synaptic connections. 
Unlike deterministic NNs, real neurons introduce noise, resulting in probabilistic rather than fixed outputs. 
This stochastic behavior helps NNs avoid local minima during training, making them more robust to noisy or incomplete data. 
However, it increases the complexity of implementation and training. Moreover, the NNs inherent randomness also reduces overfitting by preventing the exact memorization of noisy training data.

\begin{figure}[htbp]
\centering
\setlength{\unitlength}{0.8cm}
\includegraphics[width=0.7\textwidth]{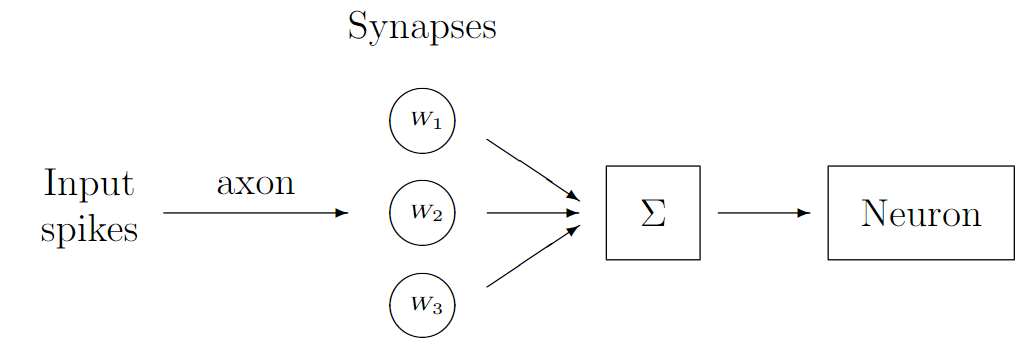}
\caption{Schematic representation of a basic artificial Neuron 
block.}\label{figure18}
\end{figure}

Stochastic neural networks consist of interconnected neurons across layers. 
A key component is the stochastic neuron, which is often implemented using \textit{magnetic tunnel junctions} (MTJs). 
An MTJ consists of two ferromagnets separated by a thin insulator and exhibits probabilistic switching behavior, see \cite{Zhu06} for details. % (see Zhu and Park~\cite{Zhu06} for details). 
In neural network models, input spikes enter weighted synapses, as Figure~\ref{figure18} illustrates.

%%%%%%%%%%%%%%%%%% old version
\begin{comment}
Neurons, the basic computational units in the brain, are known to form neural networks (NNs) by connecting through synaptic junctions. 
However, real neurons introduce noise that makes the output a probabilistic function of the input. Unlike deterministic NNs, the activation map is stochastic. 
These networks are better at avoiding local minima during training. 
They are better suited to tasks with noisy or incomplete data, although implementation and training complexity increase in the stochastic case. 
In particular, their randomness improves their robustness against overfitting since they do not learn the noise in the training data precisely as deterministic networks do. 

The interconnection of neurons from different layers creates a neural network. 
The primary components for the synapses and neurons that make up a stochastic neural network are the stochastic neurons implemented by \textit{magnetic tunnel junctions} (MTJs). 
Without going into the physical details, it can be described that an MTJ device consists of two ferromagnets separated by a thin insulator and is characterized by a switching (activation) probability (see Zhu and Park~\cite{Zhu06} for more technical details). 
In the artificial neural network model, input spikes flow into synapses, each assigned a weight, as shown in Figure~\ref{figure18}.
\end{comment}

%%%%%%%%%%%%%%%%%%%%%%%%%%%%%%%%%%%%%%%

\begin{equation*}
    \Sigma:=\text{Input}\times\text{Weights}=\text{Summed Output}
\end{equation*}
defines the weighted sum, after which the neuron fires based on a threshold or activation function.

Vreeken~\cite{Vreeken03} introduced \textit{Spiking Neural Networks} (SNNs), which closely mimic biological neurons. 
Neurons emit brief electrical spikes when sufficient input is accumulated. 
These spikes travel through axons and across synapses, comprising the axon terminal, synaptic gap, and dendrite, and convey information via spatial and temporal spike patterns.
Yu et al.~\cite{Yu21} proposed the \textit{Simple and Effective Stochastic Neural Network} (SE-SNN), which models activation uncertainty at each layer by predicting a Gaussian mean and variance and sampling during the forward pass.
This approach improves robustness in pruning, adversarial defense, learning with label noise, and model calibration with the help of an activation regularizer.

Rather than binary encoding, values can be represented by the probability of encountering ones in bit streams, 
which is central to stochastic computing (SC) \cite{Brown01}. 
SC uses random bit streams and standard digital logic for computation, allowing for simpler, fault-tolerant hardware. 
Functions such as sigmoid and tanh are implemented via linear finite state machines, reducing cost at some precision's expense. 
As noted in \cite{Liu20}, the noise in SC can also reduce overfitting and enhance inference accuracy. 
Both inputs and outputs are represented as probabilistic bit streams, merging stochastic methods with digital computation in a novel data processing framework.

\section*{Acknowledgments}
% The authors would like to thank the referees for the helpful suggestions.
%
Matthias Ehrhardt acknowledges funding by the Deutsche Forschungsgemeinschaft (DFG, German Research Foundation) -- Project-ID 531152215 -- CRC 1701.

%%%%%%%%%%%%%%%%%%%%%%%%%%%%%%%%
\newpage
\subsection*{Some Remarks and Suggestions}\label{subsec:remarks_suggestions}
{\bf On the Validity of Proposition~3.1 and the Role of Coisotropy\\}
In Proposition~3.1, the stochastic energy balance for an implicit generalized stochastic port-Hamiltonian system is formulated via Stratonovich integration:
\begin{equation}
    H(X_t) - H(X_0) = \int_0^t \langle \nabla H(X_s), \delta X_s \rangle,
    \label{eq:pathwise_energy}
\end{equation}
while this pathwise identity is formally correct in the framework of Stratonovich calculus, it implicitly assumes that the stochastic perturbations are compatible with the geometry of the Hamiltonian, 
in the sense that they do not induce any net drift in the energy evolution. 
To rigorously ensure the validity of the stochastic energy balance \eqref{eq:pathwise_energy}, we must verify that the stochastic dynamics satisfy a \emph{coisotropy condition}:
\begin{equation}
    \nabla H(x)^\top \Sigma(x) = 0 \quad \text{for all } x \in \mathcal{X},
    \label{eq:coisotropy}
\end{equation}
where $\Sigma$ is the diffusion matrix of the It\^{o} representation of the system, hence asserting that the stochastic perturbations act tangentially to the level sets of the Hamiltonian.
Therefore, any diffusion-driven motion does not change the energy of the system. 
Geometrically, the diffusion vector fields must lie in the coisotropic subspace orthogonal to $\nabla H$, thereby preserving the energy structure of the Hamiltonian dynamics. 
Without this assumption \eqref{eq:coisotropy}, the stochastic integral in \eqref{eq:pathwise_energy} may possess a non-zero drift. Therefore the identity may not hold in expectation or even in distribution. 
This issue becomes particularly relevant when attempting to apply probabilistic tools, as in the case of Dynkin's formula, which relies (also) on It\^{o} calculus, hence requiring knowledge of the drift and diffusion terms of the process.

To overcome these limitations of the pathwise formulation and to ensure compatibility with probabilistic analysis, we reformulate the energy balance in the weak (expectation) sense using the infinitesimal generator of the underlying It\^{o} diffusion. 

Let us consider a stochastic port-Hamiltonian system (SPHS) evolving under the It\^{o} SDE:
\begin{equation}
    dX_t = f(X_t) \, dt + \Sigma(X_t) \, dW_t,
    \label{eq:ito_dynamics}
\end{equation}
where % ( let me recap, just for the sake of my understanding :) ) 
$X_t \in \mathbb{R}^n$ is the state vector, $f\colon\mathbb{R}^n \to \mathbb{R}^n$ is the drift vector field, $\Sigma\colon\mathbb{R}^n \to \mathbb{R}^{n \times m}$ is the diffusion matrix, and $W_t$ is an $m$-dimensional Brownian motion adapted to the filtration $(\mathcal{F}_t)_{t \ge 0}$.
In the port-Hamiltonian formulation, the drift vector $f$ has the structure:
\begin{equation}
    f(x) = \bigl(J(x) - R(x)\bigr) \nabla H(x) + G(x) u(t),
    \label{eq:phs_drift}
\end{equation}
where (as before):
\begin{itemize}
    \item $H\colon\mathbb{R}^n \to \mathbb{R}$ is the Hamiltonian, i.e., the energy function,
    \item $J(x)$ is a skew-symmetric structure matrix representing the interconnection (satisfying $J(x)^\top = -J(x)$),
    \item $R(x)$ is a symmetric positive semi-definite matrix modeling dissipation,
    \item $G(x)$ is the control port matrix,
    \item $u(t)$ is the control input.
\end{itemize}
Accordingly, we compute the infinitesimal generator $\mathcal{L}$ of the diffusion process $X_t$ applied to the observable $H$.
Since $H \in C^2$, the generator reads:
\begin{equation}
    \mathcal{L}H(x) = \nabla H(x)^\top f(x) 
    + \frac{1}{2} \operatorname{tr} \bigl( \Sigma(x) \Sigma(x)^\top \nabla^2 H(x) \bigr),
    \label{eq:ito_generator}
\end{equation}
and, under the coisotropy condition \eqref{eq:coisotropy}, the first term $\nabla H^\top \Sigma$ vanishes identically. 
Consequently, the diffusion term in \eqref{eq:ito_generator} does not contribute to the drift of $H(X_t)$ in the It\^{o} expansion.
Therefore, we have:
\begin{equation}
    \mathcal{L} H(x) = \nabla H(x)^\top \bigl(J(x) - R(x)\bigr) \nabla H(x) + \nabla H(x)^\top G(x) u(t),
\end{equation}
and via the skew-symmetry of $J$, we observe that:
\begin{equation*}
   \nabla H(x)^\top J(x) \nabla H(x) = 0,
\end{equation*}
and hence:
\begin{equation}
    \mathcal{L} H(x) = - \nabla H(x)^\top R(x) \nabla H(x) + y(x)^\top u(t),
    \label{eq:gen_H_with_y}
\end{equation}
where $y(x) = G(x)^\top \nabla H(x)$ is the output port variable conjugate to the input $u(t)$.

Now, applying Dynkin's formula to the process $H(X_t)$, we obtain the \textit{weak energy balance}:
\begin{equation}
    \frac{d}{dt} \mathbb{E}\bigl[H(X_t)\bigr] 
    = \mathbb{E}\bigl[\mathcal{L}H(X_t)\bigr] 
    = - \mathbb{E}\bigl[\nabla H(X_t)^\top R(X_t) \nabla H(X_t)\bigl]
     + \mathbb{E}\bigl[y_t^\top u_t\bigr].
    \label{eq:weak_energy_balance}
\end{equation}
Moreover, if $R(x) \succeq 0$ for all $x$, then the first term in \eqref{eq:weak_energy_balance} is non-positive, and we obtain the inequality:
\begin{equation}
    \frac{d}{dt} \mathbb{E}\bigl[H(X_t)\bigr] 
    \le \mathbb{E}\bigl[y_t^\top u_t\bigr],
    \label{eq:passivity_inequality}
\end{equation}
which expresses the \textit{weak passivity} of the system in the sense of expected energy.
Let us underline that
%\subsection*{Discussion and Implications}
Equation \eqref{eq:passivity_inequality} provides a rigorous foundation for the expected energy dissipation of SPHSs %stochastic port-Hamiltonian systems
and constitutes a more robust and mathematically sound alternative to the pathwise identity \eqref{eq:pathwise_energy}: 
it shows that energy dissipation in expectation is guaranteed under two verifiable \textit{structural assumptions}:
\begin{enumerate}
    \item The damping matrix $R(x)$ is positive semi-definite;
    \item The coisotropy condition \eqref{eq:coisotropy}: $\nabla H^\top \Sigma = 0$ holds globally.
\end{enumerate}
These hypotheses ensure that the stochastic perturbations do not inject energy into the system on average, and that the deterministic dissipation encoded in $R(x)$ governs the energy decay.

\begin{remark}
    We also remark that the passivity inequality \eqref{eq:passivity_inequality} extends naturally to more general settings, including time-varying or control-dependent Hamiltonians, provided suitable regularity and compatibility conditions are satisfied. 
Furthermore, this framework enables a natural definition of a \emph{storage function} and the use of stochastic Lyapunov techniques to assess stability in expectation.
\end{remark}

Accordingly to what above, we could think about addressing the 
missing verification of coisotropy \eqref{eq:coisotropy} in Proposition~3.1, providing the following amendments to the paper:
\begin{itemize}
    \item Introduce the coisotropy condition \eqref{eq:coisotropy} explicitly as a structural assumption required for the validity of the energy balance \eqref{eq:passivity_inequality}.
    \item Add a new proposition (as above) that derives the energy balance in the weak sense using the It\^{o} generator and Dynkin's formula.
    \item Optionally, retain Proposition~3.1 as a formal identity under the Stratonovich calculus, but clearly delimit its domain of validity and contrast it with the expectation-based result.
\end{itemize}
\begin{comment}
\Tmatthias{We can address all 3 issues. I would keep Proposition~3.1 as it is. Afterwards, put a remark  on the limits of its domain of validity and doing so, motivate the expectation-based result next, i.e.\ introduce the coisotropy condition \eqref{eq:coisotropy} explicitly as a structural assumption required for the validity of the energy balance \eqref{eq:passivity_inequality}. Finally, end up with the formulation of the corresponding (new) proposition  that derives the energy balance in the weak sense  \eqref{eq:passivity_inequality} using the It\^{o} generator and Dynkin’s formula.}
\end{comment}

What is above preserves the original Stratonovich formulation's geometric intuition, enhances the paper's analytic rigor, and bridges the gap with stochastic control theory. 
Moreover, we align the generalized stochastic Dirac structures introduced earlier in the manuscript, while ensuring consistency across the geometric and probabilistic formulations.

\medskip

\paragraph{Point (5) - Passivity in the Stochastic Regime (detailed).}
%More for my own peace of mind than for the sake of explaining something to anyone, let me start by 
Let us start by reminding the reader that strong passivity is declared through the pathwise inequality
\begin{equation*}
   H(X_t) \le H(X_0) + \int_0^t u^\top(s)\,y(s)\,\delta Z^C_s,
\tag{P$_{\mathrm{s}}$}
\end{equation*}
where the storage is the Hamiltonian $H$ and the supply rate is $u^\top y$.  
For the  stochastic PHS
\begin{equation*}
  \delta X_t = \bigl(\mathbf{J} - \mathbf{R}\bigr) \nabla H(X_t) \,\delta Z_t
  +\mathbf{G}(X_t) u_t\,\delta Z^C_t 
  +\Sigma(X_t)\,\delta W_t,
\end{equation*}
the ensuing energy balance \eqref{calc} contains the Stratonovich line integral
$\int_0^t\nabla H^\top \mathbf{R}\,\nabla H\,\delta Z_s$.
Imposing its \emph{pathwise} non-negativity (condition \eqref{adcond}) is generically impossible, even with $\mathbf{R}\ge0$, because the signed measure~$\delta Z_s$ may reverse sign along a single trajectory.
Moreover, concerning the \textit{Mean-Square Passivity} (MSP), 
replace (P$_{\mathrm{s}}$) by the \emph{mean-square} dissipativity inequality
\begin{equation*}
\quad
\mathbb{E}\Bigl[H(X_t)\Bigr] - \mathbb{E}\Bigl[H(X_0)\Bigr] \le \mathbb{E}\left[\int_0^t u^\top(s)\,y(s)\,ds\right]\quad
\tag{P$_{\mathrm{ms}}$}
\end{equation*}
together with the \emph{coisotropy constraint}
\begin{equation*}
  \nabla H(x)^\top\Sigma(x)=0,\qquad\forall x\in X,
\tag{C$_\Sigma$}
\end{equation*}
which forces the diffusion to act tangentially to the energy level sets. And we can state a result of the following type
%%%%%%%%%%%%%%%%%%%%%%%%%%%%%
\begin{theorem}[MSP Sufficiency]
Assume $H\in C^2$, $\mathbf{R}(x)\succeq 0$, and (C$_\Sigma$).  
Let $\mathcal{L}$ be the It\^{o} generator
$\mathcal{L}f=\tfrac{1}{2}\operatorname{tr}\bigl(\Sigma\Sigma^\top\nabla^2 f\bigr) + \bigl[(\mathbf{J}-\mathbf{R})\nabla H+\mathbf{G} u\bigr]^\top\nabla f$.
Then
\begin{equation*}
\mathcal{L}H = -\nabla H^\top \mathbf{R}\,\nabla H\le u^\top y,
\end{equation*}
and consequently inequality~\emph{(P\(_{\mathrm{ms}}\))} holds for all $t\ge0$.
\end{theorem}

\emph{Sketch of a possible proof}
Apply Dynkin's formula to $H(X_t)$ and exploit (C$_\Sigma$) to cancel the diffusion term; since $\mathbf{J}$ is skew, $\nabla H^\top\mathbf{J} \nabla H\equiv0$, and the remainder yields the stated bound.

\medskip
\textbf{Storage Functions \& Martingale Structure.}
Under MSP, $V(t):=H(X_t) - \int_0^t u^\top y\,ds$ is a supermartingale:
$\mathbb{E}[V(t)]\le\mathbb{E}[V(0)]$,
hence $H$ acts as a Lyapunov function \textit{in expectation}; stability follows by Doob's supermartingale convergence theorem.
Concerning what we can define as the problem of 
designing  guidelines for Data-Driven models, we have [I know what follows is very sketchy, but we deepen the arguments, in case we decide to harm ourselves :) ]
\begin{enumerate}
%[label=\roman*),leftmargin=*,itemsep=2pt]
\item \textit{Positive damping.}  
Parameterize $\mathbf{R}_\theta = \mathbf{L}_\theta\mathbf{L}_\theta^\top$ (Cholesky or softplus), guaranteeing $\mathbf{R}_\theta\succeq0$ during training.
\item \emph{Noise orthogonality.}  Constrain $\Sigma_\theta$ via a projection
$\Sigma_\theta(x):=\bigl(\mathbf{I} - \tfrac{\nabla H_\theta\nabla H_\theta^\top}{\|\nabla H_\theta\|^2}\bigr)\,\widetilde{\Sigma}_\theta(x)$,
which enforces (C$_\Sigma$).
\item \emph{Loss augmentation.}  
Add $\lambda\bigl\lVert(\nabla H_\theta)^\top \Sigma_\theta\bigr \rVert_2^2$
to the training loss and fit $\lambda\gg0$ adaptively.
\end{enumerate}

\medskip
\textbf{Interconnection.}
For two MSP systems $(\Sigma_1,\mathbf{R}_1,H_1)$ and $(\Sigma_2,\mathbf{R}_2,H_2)$ interconnected via power ports, the composite storage $H:=H_1+H_2$
remains MSP, because the supermartingale property is additive and the supply rates cancel internally.  
This generalizes the deterministic passivity theorem to the stochastic domain.

\medskip
\textbf{Final considerations}\\
% Again, please forgive my verbosity, but I 
Let us emphasize that strong (pathwise) passivity cannot be upheld for generic diffusion-driven port-Hamiltonian systems.
Consider any semimartingale $Z_t$ with unbounded variation, such as Brownian motion or a L\'evy process. 
The Stratonovich integral, which appears in the energy balance, is given by 
$I = \int_0^t \nabla H^\top R\,\nabla H\,\delta Z_s$.
Even when $\mathbf{R}\succeq 0$, the integrand is nonnegative, while the differential $\delta Z_s$ changes sign infinitely often on almost every trajectory. 
Consequently, the integral itself oscillates and may be positive or negative on arbitrarily short intervals.  
A pointwise inequality of the form
\begin{equation*}
    H(X_t)\le H(X_0) + \int_0^t u^\top(s)\,y(s)\,\delta Z^C_s
\end{equation*}
is therefore violated with probability one, unless $Z_t$ has bounded variation. 
This justifies abandoning strong passivity in favor of a weaker, yet attainable, notion.
Moreover, \textit{mean-square passivity} (MSP) replaces the pathwise energy constraint with an expectation inequality
\begin{equation*}
    \mathbb{E}\bigl[H(X_t)\bigr]-\mathbb{E}\bigl[H(X_0)\bigr]\le
    \mathbb{E}\Bigl[\int_0^t u^\top(s)\,y(s)\,ds\Bigr].
\tag{MSP}
\end{equation*}
This inequality (MSP) states that the \textit{expected} stored energy never exceeds the supplied work.
Indeed, by writing the It\^{o} dynamics in the form $dX_t=f(X_t,u_t)\,dt+\Sigma(X_t)\,dW_t$ with $f=(\mathbf{J} - \mathbf{R})\nabla H+\mathbf{G} u$ and imposing the orthogonality constraint $\nabla H^\top \Sigma\equiv 0$, the It\^{o} generator becomes  
$\mathcal{L} H = -\nabla H^\top \mathbf{R}\,\nabla H + u^\top y$, and since $\mathbf{R}$ is positive semidefinite, $\mathcal{L} H\le u^\top y$, and Dynkin's formula yields (MSP) automatically. 
Hence the requirement \textit{boils down} to two constructive conditions: positivity of the damping matrix $\mathbf{R}$ and covariance-orthogonality of the diffusion $\Sigma$ concerning the Hamiltonian gradient. 
Such conditions can be encoded directly in data-driven (neural) surrogates.  
A common strategy is to parameterize the damping through a Cholesky factor, $\mathbf{R}_\theta=L_\theta L_\theta^\top$, guaranteeing $\mathbf{R}_\theta\succeq 0$ for all network parameters.  
Orthogonality of the diffusion can be enforced by projecting a free network output $\widetilde{\Sigma}_\theta$ onto the tangent space of the energy level sets,
\begin{equation*}
  \Sigma_\theta(x)=
   \Bigl(\mathbf{I} - \frac{\nabla H_\theta\nabla H_\theta^\top}{\|\nabla H_\theta\|_2^{2}+\varepsilon}\Bigr)\,\widetilde{\Sigma}_\theta(x),
\end{equation*}
where $\varepsilon$ is a small constant preventing numerical blow-up when $\nabla H_\theta$ is near zero. 
In practice, one augments the training loss with large penalties on both the violation of the orthogonality condition and the discrepancy $\mathcal{L}_\theta H_\theta-u^\top y_\theta$.  
After training, long-horizon roll-outs can be used to verify empirically that the map $t\mapsto\mathbb{E}[H_\theta(X_t)]-\mathbb{E}\bigl[\int_0^t u^\top y_\theta\,ds\bigr]$  
is non-increasing, confirming that MSP holds in simulation.

Unlike the unattainable strong passivity constraint, MSP is achievable for diffusion-driven systems, retains the intuitive \textit{energy cannot be created on average} interpretation, remains closed under interconnection of subsystems (the sum of two MSP storage functions is again MSP), a perfect (well done) puzzle with statistical learning where expectations are estimated empirically.  
Summing up, unless we want to keep chasing unicorns in the Stratonovich jungle, we might as well accept that MSP isn't just a convenient compromise; it's the only passivity notion that won't collapse under the weight of actual stochasticity. % Let's call it what it is: the grown-up way to do passivity in the neural port-Hamiltonian world :)

%%%% Bibliography  %%%%%%%%%% alphabetic order

\end{document}